\documentclass{mn2e}
\usepackage{graphicx}

\begin{document}

\title[RCSLenS: gravitational physics through
  cross-correlation]{RCSLenS: Testing gravitational physics through
  the cross-correlation of weak lensing and large-scale structure}

\author[Blake et al.]{\parbox[t]{\textwidth}{Chris
    Blake$^1$\footnotemark, Shahab Joudaki$^1$, Catherine Heymans$^2$,
    Ami Choi$^2$, Thomas \\ Erben$^3$, Joachim Harnois-Deraps$^4$,
    Hendrik Hildebrandt$^3$, Benjamin Joachimi$^5$, \\ Reiko
    Nakajima$^3$, Ludovic van Waerbeke$^4$ and Massimo Viola$^6$}
  \\ \\ $^1$ Centre for Astrophysics \& Supercomputing, Swinburne
  University of Technology, P.O.\ Box 218, Hawthorn, VIC 3122,
  Australia \\ $^2$ Scottish Universities Physics Alliance, Institute
  for Astronomy, University of Edinburgh, Royal Observatory, Blackford
  Hill, \\ Edinburgh, EH9 3HJ, U.K. \\ $^3$ Argelander Institute for
  Astronomy, University of Bonn, Auf dem Hugel 71, 53121, Bonn,
  Germany \\ $^4$ Department of Physics and Astronomy, University of
  British Columbia, 6224 Agricultural Road, Vancouver, V6T 1Z1, B.C.,
  Canada \\ $^5$ Department of Physics and Astronomy, University
  College London, London WC1E 6BT, U.K. \\ $^6$ Leiden Observatory,
  Leiden University, Niels Bohrweg 2, 2333 CA Leiden, The Netherlands}

\maketitle

\begin{abstract}
The unknown nature of ``dark energy'' motivates continued cosmological
tests of large-scale gravitational physics.  We present a new
consistency check based on the relative amplitude of non-relativistic
galaxy peculiar motions, measured via redshift-space distortion, and
the relativistic deflection of light by those same galaxies traced by
galaxy-galaxy lensing.  We take advantage of the latest generation of
deep, overlapping imaging and spectroscopic datasets, combining the
Red Cluster Sequence Lensing Survey (RCSLenS), the
Canada-France-Hawaii Telescope Lensing Survey (CFHTLenS), the WiggleZ
Dark Energy Survey and the Baryon Oscillation Spectroscopic Survey
(BOSS).  We quantify the results using the ``gravitational slip''
statistic $E_G$, which we estimate as $0.48 \pm 0.10$ at $z=0.32$ and
$0.30 \pm 0.07$ at $z=0.57$, the latter constituting the highest
redshift at which this quantity has been determined.  These
measurements are consistent with the predictions of General
Relativity, for a perturbed Friedmann-Robertson-Walker metric in a
Universe dominated by a cosmological constant, which are $E_G = 0.41$
and $0.36$ at these respective redshifts.  The combination of
redshift-space distortion and gravitational lensing data from current
and future galaxy surveys will offer increasingly stringent tests of
fundamental cosmology.
\end{abstract}

\section{Introduction}
\renewcommand{\thefootnote}{\fnsymbol{footnote}}
\setcounter{footnote}{1}
\footnotetext{E-mail: cblake@astro.swin.edu.au}

A wide set of cosmological observations suggest that the dynamics of
the Universe are currently dominated by some form of ``dark energy'',
which in standard Friedmann-Robertson-Walker (FRW) models is
propelling an acceleration in late-time cosmic expansion (e.g.\ {\it
  Planck} collaboration 2015a, Aubourg et al.\ 2015, Betoule et
al.\ 2014).  The physical nature of dark energy is not yet understood,
and a widely-considered possibility is that the nature of gravitation
differs on large cosmological scales from the predictions of General
Relativity (GR) in an FRW metric.  As a result, a key task for current
cosmological surveys is to construct observations to test for such
departures.

Gravitational physics produces a rich variety of observable signatures
that can be used for this purpose.  The most precisely-measured signal
results from the ``peculiar motions'' of galaxies as they fall toward
overdense regions as non-relativistic test particles in a perturbed
FRW metric.  These motions produce correlated Doppler shifts in galaxy
redshifts that manifest themselves as an overall anisotropy in the
measured clustering signal as a function of the angle to the
line-of-sight (Kaiser 1987), known as redshift-space distortion (RSD).
The amplitude of this anisotropy has been accurately measured in a
number of galaxy surveys across a range of redshifts and allows the
growth rate $f$ of cosmic structure, which describes the gravitational
amplification of density perturbations, to be inferred.  To date,
these measurements are broadly consistent with the prediction of the
standard cosmological model (e.g.\ Blake et al.\ 2011, Beutler et
al.\ 2012, Reid et al.\ 2012, de la Torre et al.\ 2013, Samushia et
al.\ 2014).

A highly complementary route for probing gravitational effects is to
study the deflections of relativistic test particles such as photons,
which are additionally sensitive to the curvature of space produced by
density perturbations.  These deflections, known as gravitational
lensing, may be measured in a statistical sense using the correlations
imprinted in the apparent shapes of background galaxies behind
foreground lenses in deep imaging surveys.  The level of the signal is
determined by both the amplitude of the density fluctuations around
the lenses (again reflecting the growth of structure with redshift)
and the relative distances of the source-lens systems, both of which
may be predicted by a given cosmological model (for a review, see
Bartelmann \& Schneider 2001).  Recent projects such as the
Canada-France-Hawaii Telescope Lensing Survey (CFHTLenS) have allowed
a suite of such tests to be carried out (e.g.\ Heymans et al.\ 2013,
Simpson et al.\ 2013) by providing deep, wide, high-resolution
imaging.

Zhang et al.\ (2007) proposed that a powerful gravitational
consistency check might be performed by using the same set of galaxies
to trace non-relativistic gravitationally-driven motion using
redshift-space distortion, and to serve as foreground lenses for
probing the relativistic deflection of light from background sources.
In this way, it could be ascertained whether the relative amplitude of
these two effects, driven by the same underlying density perturbations
traced by the lenses, was consistent with the prediction of GR
assuming a perturbed FRW metric for a given set of cosmological
parameters including the matter density $\Omega_m$.  This can be
achieved by measuring a quantity known as the ``gravitational slip''
$E_G(R)$ as a function of physical scale $R$, which is constructed
from the galaxy-galaxy lensing signal and the redshift-space
distortion and clustering amplitude of the lenses.  Standard perturbed
GR cosmology predicts that a scale-independent value $E_G =
\Omega_m/f$ should be recovered.  Failure of this cross-check would
indicate either the breakdown of linear perturbation theory, an
inconsistency in the assumed cosmological parameters such as the
matter density or curvature, or that a large-scale modification in
gravitational physics was required.

A requirement for carrying out this programme is the availability of
galaxy spectroscopic redshift surveys and deep gravitational lensing
imaging surveys which overlap on the sky.  Reyes et al.\ (2010)
performed this test using data from the Sloan Digital Sky Survey
(SDSS) covering $\approx 5000$ deg$^2$, using a sample of $\approx
70{,}000$ Luminous Red Galaxy lenses at $z=0.32$ and shape
measurements with a surface density $\approx 1$ arcmin$^{-2}$ and
produced a gravitational slip measurement consistent with GR in a
$\Lambda$-dominated Universe.  The goal of our current study is to use
new, significantly deeper, spectroscopic and imaging survey datasets
to extend these tests to higher source densities and more distant
redshifts $z \approx 0.6$.  In particular we use imaging data from
CFHTLenS (Heymans et al.\ 2012) and the 2nd Red Cluster Sequence
Lensing Survey (Gilbank et al.\ 2011, Hildebrandt et al.\ 2015), and
overlapping spectroscopic data from the WiggleZ Dark Energy Survey
(Drinkwater et al.\ 2010) and Baryon Oscillation Spectroscopic Survey
(BOSS, Eisenstein et al.\ 2011) to carry out this investigation.  We
note that a measurement of $E_G$ using lensing of the Cosmic Microwave
Background was recently presented by Pullen et al.\ (2015).

The structure of our paper is as follows: in Section \ref{sectheory}
we introduce the theory underpinning galaxy-galaxy lensing and RSD in
a perturbed FRW metric, the methodology for suppressing small-scale
information which is difficult to model, and our test statistic for
gravitational physics, $E_G$.  In Section \ref{secdata} we summarize
our input datasets, and in Section \ref{secmeas} we present the
galaxy-galaxy lensing measurements expressed as both the average
tangential shear as a function of angular separation
$\gamma_t(\theta)$ and the differential surface density as a function
of projected physical separation $\Delta \Sigma(R)$, together with a
series of tests for shape-measurement systematics.  We pay particular
attention to how the full redshift probability distribution of each
source, determined from photometric-redshift estimation, can be
included in an unbiased estimator of $\Delta \Sigma(R)$ from
source-lens pairs.  In Section \ref{secsim} we describe a large set of
new N-body simulations, including self-consistent gravitational
lensing, that we use for determining errors in our measurements and
testing models.  Finally in Section \ref{seccosmo} we present the
first cosmological implications, including new measurements of $E_G$
up to redshift $z \approx 0.6$.  We conclude in Section \ref{secconc}.

This paper is the first in a series which will dissect the
cosmological information in overlapping deep galaxy lensing and
spectroscopic datasets.  Future studies will present full cosmological
fits including Cosmic Microwave Background data in a range of modified
gravity scenarios, and the use of photometric-redshift and
spectroscopic-redshift cross-correlations to determine simultaneously
the source redshift distributions and cosmological parameters.

\section{Theory: galaxy-galaxy lensing and redshift-space distortion}
\label{sectheory}

\subsection{Galaxy-galaxy lensing: tangential shear}

Galaxy-galaxy lensing describes the shear-density correlation
imprinted between the shapes of background source galaxies and the
foreground large-scale structure traced by the lens galaxies; it is
measured by computing the azimuthally-averaged tangential shear of the
sources as a function of their angular distance $\theta$ from the
lenses:
\begin{equation}
\langle \gamma_t(\theta) \rangle = \langle \delta_g(\vec{x}) \,
\gamma_t(\vec{x} + \vec{\theta}) \rangle ,
\label{eqgtdef}
\end{equation}
where $\gamma_t$ denotes the tangential shear component of the source
with respect to the separation vector $\vec{\theta}$ connecting it to
the lens, and $\delta_g(\vec{x})$ is the overdensity of the lens
galaxies at position $\vec{x}$.  Fourier transforming the variables in
Equation \ref{eqgtdef} we find that (e.g.\ Hu \& Jain 2004)
\begin{equation}
\langle \gamma_t(\theta) \rangle = \frac{1}{2\pi} \int_0^\infty d\ell
\, \ell \, C_{g\kappa}(\ell) \, J_2(\ell \theta) ,
\label{eqgtmod}
\end{equation}
where $J_2$ denotes the 2nd-order Bessel function of the first kind,
and the galaxy-convergence cross-power spectrum $C_{g\kappa}$ is given
by (e.g.\ Bartelmann \& Schneider 2001, Guzik \& Seljak 2001, Joachimi
\& Bridle 2010)
\begin{equation}
C_{g\kappa}(\ell) = \frac{3 \Omega_m H_0^2}{2 c^2} \int_0^\infty dz
\frac{(1+z)}{\chi(z)} P_{gm} \left( \frac{\ell}{\chi},z \right) p_l(z)
W(z) ,
\label{eqpgk}
\end{equation}
where
\begin{equation}
W(z) = \int_z^\infty dz' \, p_s(z') \left[ \frac{\chi(z') -
    \chi(z)}{\chi(z')} \right] .
\label{eqws}
\end{equation}
In these equations, $\chi(z)$ is the co-moving radial coordinate at
redshift $z$ assuming a spatially-flat Universe, $P_{gm}(k,z)$ is the
galaxy-mass cross-power spectrum at wavenumber $k$, $[p_s(z), p_l(z)]$
are the redshift probability distributions of the source and lens
samples, normalized such that $\int_0^\infty p(z) \, dz = 1$, and
$H_0$ and $c$ are Hubble's constant and the speed of light,
respectively.  These relations make the approximations of using the
Limber equation and neglecting additional effects such as cosmic
magnification and intrinsic alignments.

These equations make clear that the tangential shear is sensitive to
the cosmological model through the distance-redshift relation
describing the relative geometry of the source-lens-observer systems,
and through the clustering of the matter overdensities traced by the
lenses.  Our method of estimating $\langle \gamma_t(\theta) \rangle$
from the data is described in Section \ref{secgtest}.

\subsection{Galaxy-galaxy lensing: differential surface density}
\label{secdelsigtheory}

In this study we are interested in ``de-projecting'' the angular
tangential shear profile in order to recover the statistics of the
projected mass surface density $\Sigma$ around the lenses as a
function of projected spatial separation $R$.  Assuming an
axisymmetric lens, and a single lens-source pair with respective
redshifts $(z_l, z_s)$, the average tangential shear at projected
separation $R$ from the lens is given by
\begin{equation}
\langle \gamma_t(\theta) \rangle = \frac{\Delta \Sigma(R,
  z_l)}{\Sigma_c(z_l, z_s)}
\label{eqgtdelsig1}
\end{equation}
(e.g.\ Bartelmann \& Schneider 2001), where $\theta = R/\chi(z_l)$ and
the differential projected surface mass density $\Delta \Sigma$ is
defined in terms of $\Sigma(R)$ by
\begin{equation}
\Delta \Sigma(R) \equiv \overline{\Sigma}(<R) - \Sigma(R) ,
\label{eqdelsigdef}
\end{equation}
where $\overline{\Sigma}(<R)$ is the average mass density within a
circle of radius $R$,
\begin{equation}
\overline{\Sigma}(<R) \equiv \frac{2}{R^2} \int_0^R R' \, \Sigma(R')
\, dR' .
\end{equation}
In Equation \ref{eqgtdelsig1} the co-moving critical surface mass
density $\Sigma_c$ is given by
\begin{equation}
\Sigma_c(z, z') = \frac{c^2}{4 \pi G} \left\{ \frac{\chi(z')}{ \left[
    \chi(z') - \chi(z) \right] \, \chi(z) \, (1 + z)} \right\} ,
\label{eqsigc}
\end{equation}
where $G$ is the gravitational constant and $z' > z$.  The
differential projected surface density around the lenses is related to
the lens galaxy-matter cross-correlation function $\xi_{gm}(r)$ by
\begin{equation}
\Delta \Sigma(R) = \overline{\rho_m} \int_{-\infty}^\infty
\xi_{gm}(\sqrt{R^2 + \Pi^2}) \, d\Pi ,
\label{eqxigm}
\end{equation}
where $\overline{\rho_m}$ is the mean cosmological matter density,
related to the critical density $\rho_c = 3 H_0^2/8 \pi G = 2.77518
\times 10^{11} \, h^2 M_\odot$ Mpc$^{-3}$ [with $h = H_0/(100$ km
  s$^{-1}$ Mpc$^{-1}$)] by $\overline{\rho_m} = \rho_c \, \Omega_m$,
and $\Pi$ denotes the co-moving line-of-sight separation.

By substituting Equations \ref{eqpgk} and \ref{eqws} into Equation
\ref{eqgtmod}, and comparing with the result of substituting Equations
\ref{eqsigc} and \ref{eqxigm} into Equation \ref{eqgtdelsig1}, we can
determine after some algebra how Equation \ref{eqgtdelsig1}
generalizes for broad source and lens distributions.  The result is
\begin{equation}
\langle \gamma_t(\theta) \rangle = \int_0^\infty dz \, p_l(z) \,
\Delta \Sigma(R, z) \int_z^\infty dz' \, p_s(z') \frac{1}{\Sigma_c(z,
  z')} ,
\label{eqgtdelsig2}
\end{equation}
noting that for narrow lens and source redshift distributions, $p_l(z)
= \delta_D(z-z_l)$ and $p_s(z) = \delta_D(z-z_s)$, Equation
\ref{eqgtdelsig2} reduces to Equation \ref{eqgtdelsig1}.

Equation \ref{eqgtdelsig2} clarifies that for a broad lens redshift
distribution it is not possible to solve for $\Delta \Sigma(R, z_l)$
in closed form.  However, for a narrow lens redshift distribution,
averaging over a background source redshift distribution $p_s(z)$, we
can recover the relation
\begin{equation}
\Delta \Sigma(R, z_l) \approx \langle \gamma_t(\theta) \rangle \left[
  \int_{z_l}^\infty dz' \, p_s(z') \frac{1}{\Sigma_c(z_l, z')}
  \right]^{-1} .
\label{eqgtdelsig3}
\end{equation}
Section \ref{secdelsigest} presents an analysis to justify that this
approximation is acceptable, given the source and lens redshift
distributions of our datasets.  Our method of estimating $\Delta
\Sigma(R)$ from the data is described in Section \ref{secdelsigest}.

\subsection{Galaxy clustering and redshift-space distortion}

The cross-power spectrum of lens galaxies and underlying mass that
appears in Equation \ref{eqpgk}, $P_{gm}$, and the equivalent
cross-correlation function in Equation \ref{eqxigm}, $\xi_{gm}$,
depend on the manner or ``bias'' with which the lens galaxies trace
the matter field.  Although this bias is in general a stochastic,
non-linear and scale-dependent function, it may be approximated on
sufficiently-large scales as a linear mapping between the galaxy and
matter overdensity, $\delta_g(\vec{x}) = b \, \delta_m(\vec{x})$
(e.g., Scherrer \& Weinberg 1998).  In this case $\xi_{gm}(r) = b \,
\xi_{mm}(r)$, in terms of the matter auto-correlation function
$\xi_{mm}(r)$ which may be modelled from the cosmological parameters,
where the unknown bias parameter $b$ may be determined using the
auto-correlation function of the lenses, $\xi_{gg}(r) = b^2 \,
\xi_{mm}(r)$.

Since 3D measurements of galaxy clustering are performed in
redshift-space, the apparent radial positions of the galaxies contain
an additional correlated imprint from galaxy peculiar velocities,
known as redshift-space distortion (RSD).  In particular, the Fourier
transform of the redshift-space galaxy overdensity field,
$\tilde{\delta}_g^s$, is given under certain assumptions
(e.g.\ Percival \& White 2009, Blake et al.\ 2011) by
\begin{equation}
\tilde{\delta}_g^s(k,\mu) = \tilde{\delta}_g(k) - \mu^2 \,
\tilde{\theta}(k) ,
\end{equation}
where $\tilde{\theta}(k)$ is the Fourier transform of the divergence
of the peculiar velocity field in units of the co-moving Hubble
velocity and $\mu$ is the cosine of the angle of the Fourier mode to
the line-of-sight.

Assuming that the velocity field is generated under linear
perturbation theory then $\tilde{\theta}(k) = - f \,
\tilde{\delta}_m(k)$, where $f$ is the growth rate of structure,
expressible in terms of the growth factor $D(a)$ at cosmic scale
factor $a$ as $f \equiv d \ln{D}/d \ln{a}$.  The growth factor is
defined in terms of the amplitude of a single perturbation as
$\delta(a) = D(a) \, \delta(1)$.  Under the assumption of linear
galaxy bias we then obtain the standard expression for the
redshift-space galaxy power spectrum in linear theory (Kaiser 1987)
\begin{equation}
P_{gg}^s(k,\mu) = b^2 \, (1 + \beta \mu^2)^2 \, P_{mm}(k) ,
\end{equation}
where we have introduced the redshift-space distortion parameter
$\beta = f/b$, which governs the amplitude of the measured RSD.

The anisotropic imprint of RSD in galaxy clustering allows the
measurement of the gravitational growth rate and, consequently,
powerful tests of gravitational physics.  However, it also introduces
an extra amplitude factor in the relation between $\xi_{gg}$ and
$\xi_{mm}$, complicating inferences about the galaxy bias.  In order
to avoid this issue the real-space ``projected'' correlation function
$w_p(R)$, independent of RSD, can instead be constructed by
integrating the 3D galaxy correlation function $\xi_{gg}(R,\Pi)$ along
the line-of-sight:
\begin{equation}
w_p(R) = \int_{-\infty}^\infty \xi_{gg}(R,\Pi) \, d\Pi .
\label{eqwpmod}
\end{equation}
Our method of estimating $w_p(R)$ from the data is described in
Section \ref{secwpest}.  In practice the limits of Equation
\ref{eqwpmod} must be taken as large, finite values.

\subsection{Suppressing small-scale information}

Equation \ref{eqdelsigdef} demonstrates that the amplitude of $\Delta
\Sigma(R)$ depends on the surface density of matter around galaxies
across a range of smaller scales from zero to $R$.  This is
problematic from the viewpoint of fitting cosmological models to the
data since at small scales, within the halo virial radius, the
cross-correlation coefficient between the matter and galaxy
fluctuations is a complex function which is difficult to predict from
first principles (Baldauf et al.\ 2010, Mandelbaum et al.\ 2010).  In
order to remove this sensitivity to small-scale information these
authors proposed a new statistic, the annular differential surface
density (ADSD), denoted by $\Upsilon$ and defined by
\begin{eqnarray}
\Upsilon_{gm}(R,R_0) &=& \Delta \Sigma(R) - \frac{R_0^2}{R^2} \,
\Delta \Sigma(R_0) \nonumber \\ &=& \frac{2}{R^2} \int_{R_0}^R R' \,
\Sigma(R') \, dR' \nonumber \\ &-& \Sigma(R) + \frac{R_0^2}{R^2}
\Sigma(R_0) ,
\label{equpsgm}
\end{eqnarray}
which does not contain information originating from scales $R < R_0$.
The small-scale limit $R_0$ is chosen to be large enough to reduce the
main systematic effects discussed above, but small enough to preserve
a high signal-to-noise ratio in the measurements (also see the
discussion in Mandelbaum et al.\ 2013).  An alternative approach is to
model the halo occupation statistics and marginalize over the free
parameters (e.g., Cacciato et al.\ 2013).

The corresponding quantity suppressing the small-scale contribution to
the galaxy auto-correlations is
\begin{eqnarray}
& & \Upsilon_{gg}(R,R_0) = \rho_c \nonumber \\ & & \left[
    \frac{2}{R^2} \int_{R_0}^R R' \, w_p(R') \, dR' - w_p(R) +
    \frac{R_0^2}{R^2} w_p(R_0) \right] .
\label{equpsgg}
\end{eqnarray}
We assume $R_0 = 1.5 \, h^{-1}$ Mpc as our fiducial value following
Reyes et al.\ (2010), and demonstrate in Section \ref{seceg} that our
results are insensitive to this choice.

\subsection{Testing gravitational physics: the $E_G$ statistic}

In general scalar theories of gravity, the perturbed FRW spacetime
metric $ds^2$ may be expressed in terms of the Newtonian potential
$\Psi$ and curvature potential $\Phi$:
\begin{equation}
ds^2 = \left[ 1 + 2 \, \Psi(\vec{x},t) \right] \, c^2 \, dt^2 - a(t)^2
\left[ 1 - 2 \, \Phi(\vec{x},t) \right] \, d\vec{x}^2 .
\end{equation}
Relativistic particles, such as photons experiencing gravitational
lensing, collect equal contributions from these two potentials as they
traverse spacetime, such that their equations of motion (and hence the
resulting lensing patterns) are determined by $\nabla^2(\Psi + \Phi)$.
However, the motion of non-relativistic particles arising from the
gravitational attraction of matter, which produces galaxy clustering
and RSD, is sensitive only to the derivatives of the Newtonian
potential $\nabla^2 \Psi$ (e.g., Jain \& Zhang 2008).

In standard General Relativity (GR), in the absence of anisotropic
stress, $\Psi(\vec{x},t) = \Phi(\vec{x},t)$ and both potentials are
related to the matter overdensity via the Poisson equation $\nabla^2
\Phi = 4 \pi G a^2 \overline{\rho_m} \delta_m$.  Therefore, by
measuring if both the gravitational lensing of photons and galaxy
peculiar velocity respond in an identical manner to the matter
overdensity traced by the lens galaxies in our datasets, we can
perform a fundamental test of whether the relation between
$(\Psi+\Phi)$ and $\Psi$ follows the GR expectation (assuming this
perturbation approximation applies).

Zhang et al.\ (2007) proposed that this test can be efficiently
carried out by cross-correlating lens galaxies to both the surrounding
velocity field using RSD and to the shear of background galaxies using
galaxy-galaxy lensing.  In particular, Reyes et al.\ (2010)
implemented this consistency test by constructing the ``gravitational
slip'' statistic
\begin{equation}
E_G(R) = \frac{1}{\beta} \frac{\Upsilon_{gm}(R,R_0)}{\Upsilon_{gg}(R,R_0)} ,
\label{eqeg}
\end{equation}
which is independent of both the galaxy bias factor $b$ and the
underlying amplitude of matter clustering $\sigma_8$, given that
$\beta \propto 1/b$, $\Upsilon_{gm} \propto b \, \sigma_8^2$ and
$\Upsilon_{gg} \propto b^2 \, \sigma_8^2$.  The perturbed GR model
prediction on large scales is then a scale-independent quantity $E_G =
\Omega_m/f$ (see Leonard et al.\ (2015) for a more detailed discussion
of this approximation).  We measure $E_G$ and carry out this
consistency test in Section \ref{seceg}.  We note that a failure of
this consistency check does not necessarily indicate evidence for
gravitational physics beyond GR: other possible explanations would
include a breakdown in validity of linear perturbation theory, or that
the value of $\Omega_m$ or curvature differs from that predicted by
measurements of the Cosmic Microwave Background radiation.

\section{Data}
\label{secdata}

\begin{figure*}
\begin{center}
\resizebox{16cm}{!}{\rotatebox{270}{\includegraphics{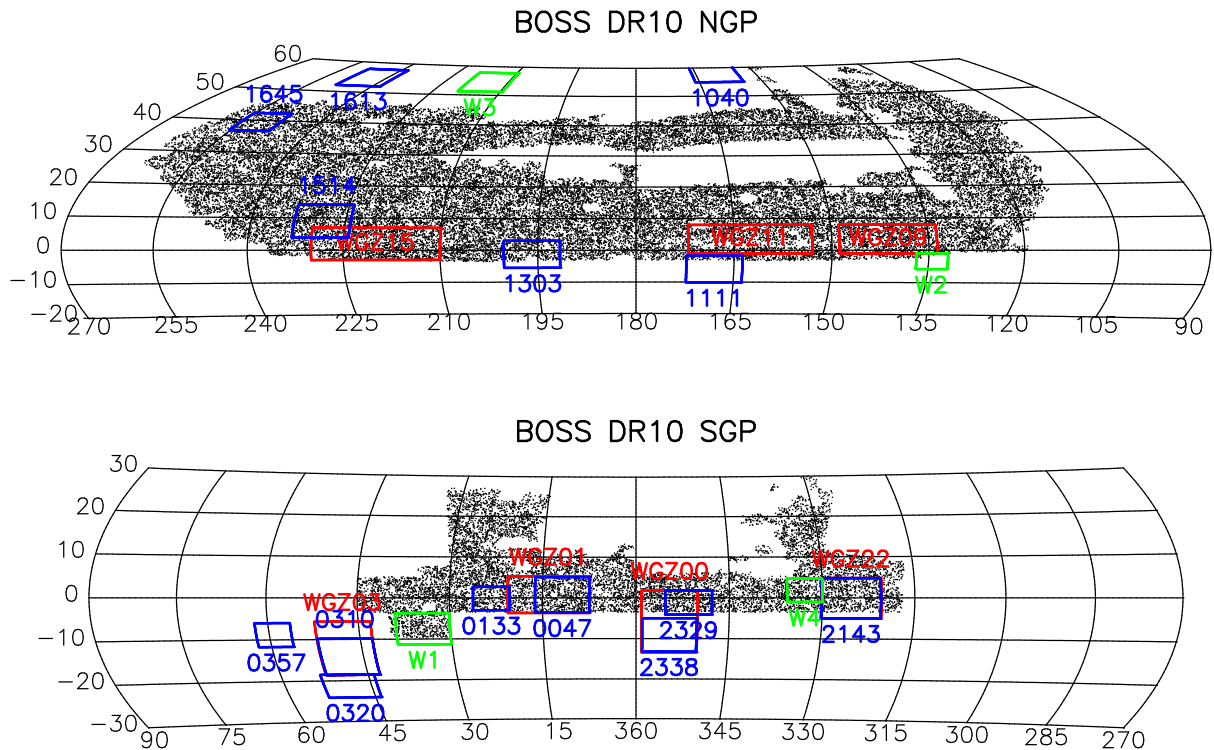}}}
\end{center}
\caption{(R.A., Dec.) distribution in the Northern Galactic Pole (NGP)
  and Southern Galactic Pole (SGP) of the datasets used in this
  analysis: the BOSS DR10 galaxy sample (black dots), the WiggleZ
  survey regions (red rectangles), the CFHTLenS fields (green
  rectangles) and the RCSLenS fields (blue rectangles).}
\label{figoverlap}
\end{figure*}

\begin{table*}
\caption{Statistics for CFHTLenS and RCSLenS fields cross-correlated
  with WiggleZ and BOSS data.  The effective (unmasked) areas are
  shown for the full source field, the pointings which contain a
  minimum of four filters such that photometric redshifts are
  available, and the pointings which overlap with the lens
  distribution (where two values are quoted for this latter area, they
  refer to WiggleZ/BOSS).  The effective weighted source density as
  computed by Equation \ref{eqsourcedens} is listed as $n_{\rm eff}$.
  The number of lenses contained in the overlapping areas of WGZLoZ,
  WGZHiZ, LOWZ and CMASS are displayed as $N_{\rm WGZLoZ}$, $N_{\rm
    WGZHiZ}$, $N_{\rm LOWZ}$ and $N_{\rm CMASS}$, respectively.  Some
  fields have overlap with both WiggleZ and BOSS lens samples, which
  would result in potentially correlated measurements.  In such cases
  the lens sample producing the lower signal-to-noise measurement,
  indicated with an asterisk, is excluded from the analysis.}
\begin{center}
\begin{tabular}{ccccccccc}
\hline
Field & $A_{\rm eff}$[all] & $A_{\rm eff}$[photo-$z$] & $A_{\rm eff}$[overlap] & $n_{\rm eff}$ & $N_{\rm WGZLoZ}$ & $N_{\rm WGZHiZ}$ & $N_{\rm LOWZ}$ & $N_{\rm CMASS}$ \\
& [deg$^2$] & [deg$^2$] & [deg$^2$] & [arcmin$^{-2}$] & & & & \\
\hline
CFHTLS W1 & 56.0 & 56.0 & 48.7 & 14.0 & - & - & 1998 & 3856 \\
CFHTLS W4 & 17.6 & 17.6 & 17.6 & 13.1 & - & - & 832 & 1711 \\
RCS 0047 & 51.8 & 37.2 & 37.2 & 5.4 & 3029$^*$ & 4343$^*$ & 2273 & 3735 \\
RCS 0133 & 25.0 & 13.3 & 13.3 & 4.5 & - & - & 646 & 1236 \\
RCS 0310 & 60.9 & 52.3 & 52.3 & 4.9 & 4249 & 6140 & - & - \\
RCS 1303 & 12.9 & 8.3 & 3.7 & 5.3 & - & - & 128 & 412 \\
RCS 1514 & 60.0 & 30.4 & 5.8/30.4 & 5.7 & 152$^*$ & 448$^*$ & 1504 & 3337 \\
RCS 1645 & 22.8 & 20.1 & 20.1 & 6.7 & - & - & 1098 & 2008 \\
RCS 2143 & 65.5 & 47.1 & 47.1 & 5.7 & 7190$^*$ & 8752$^*$ & 2337 & 4718 \\
RCS 2329 & 36.0 & 32.1 & 19.0/32.1 & 6.4 & 494$^*$ & 882$^*$ & 1356 & 3122 \\
RCS 2338 & 57.3 & 19.5 & 19.5 & 4.9 & 1336 & 2158 & - & - \\
\hline
\end{tabular}
\end{center}
\label{tabdata}
\end{table*}

We perform this test of gravitational physics by utilizing the overlap
of lensing measurements from two imaging surveys, the
Canada-France-Hawaii Telescope Lensing Survey (CFHTLenS; Heymans et
al.\ 2012) and the Red Cluster Sequence Lensing Survey (RCSLenS,
Hildebrandt et al.\ 2015), with two spectroscopic-redshift large-scale
structure surveys, the WiggleZ Dark Energy Survey (Drinkwater et
al.\ 2010) and the Baryon Oscillation Spectroscopic Survey (BOSS,
Eisenstein et al.\ 2011).  Figure \ref{figoverlap} displays the sky
distribution of the CFHTLenS, RCSLenS, WiggleZ and BOSS datasets used
in this analysis, and the surveys and source selection are briefly
described in the sub-sections below.

A total of 11 CFHTLenS and RCSLenS survey fields overlap with the
WiggleZ and BOSS DR10 data, comprising a total area of 466 deg$^2$ (74
deg$^2$ for CFHTLenS and 392 deg$^2$ for RCSLenS).  Table
\ref{tabdata} lists some statistics for these fields, including the
total effective (unmasked) field area, the area for which the
available imaging allows photometric redshifts to be derived, the
subset of that area which overlaps with the lens distributions, the
effective source density (defined below) and the number of lenses in
each of the overlapping spectroscopic surveys used in the analysis,
where the BOSS data is split into the LOWZ and CMASS samples
(described below).  The RCSLenS fields used for cross-correlation with
(WiggleZ, BOSS) contain an overlapping photo-$z$ area of (72, 184)
deg$^2$.  The CFHTLenS fields used for cross-correlation with BOSS
cover 74 deg$^2$.

Our datasets enable us to construct five distinct source-lens survey
pairings: RCSLenS-WiggleZ, RCSLenS-LOWZ, RCSLenS-CMASS, CFHTLenS-LOWZ
and CFHTLenS-CMASS.  We split the WiggleZ lenses into two independent
redshift bins, $0.15 < z < 0.43$ (``WGZLoZ'') and $0.43 < z < 0.7$
(``WGZHiZ''), which coincide with the redshift ranges of LOWZ and
CMASS, respectively, producing a total of six possible pairings.  In
the analyses that follow we will often present separate measurements
for these six cases, optimally combining the measurements in each
individual field using inverse-variance weighting.

\subsection{CFHTLenS}

The CFHTLenS\footnote{\tt http://www.cfhtlens.org} is a deep
multi-colour survey optimized for weak lensing analyses, observed as
part of the Canada-France-Hawaii Telescope (CFHT) Legacy Survey in
five optical bands $u^* g' r' i' z'$ using the 1 deg$^2$ camera
MegaCam.  The data span four fields, two of which (W1 and W4) overlap
with the spectroscopic-redshift data used in this analysis.  The main
lensing analysis is performed on $i'$ band data, for which the
observations have a $5\sigma$ point-source limiting magnitude $i'
\approx 25.5$.  The imaging data are processed with {\small THELI}
(Erben et al.\ 2013), galaxy ellipticities are measured by the
Bayesian model-fitting software {\it lens}fit (Miller et al.\ 2013),
and photometric redshifts are derived from PSF-matched photometry
(Hildebrandt et al.\ 2012) using the Bayesian photometric redshift
code {\small BPZ} (Benitez 2000).  Additive and multiplicative shear
calibration corrections have been derived as a function of galaxy size
and signal-to-noise (Heymans et al.\ 2012, Miller et al.\ 2013).  The
survey pointings have been subjected to a stringent
cosmology-independent systematic-error analysis (Heymans et
al.\ 2012), as a result of which a subset of around $25\%$ of the
pointings have been flagged as possessing potential systematic errors.
Given that galaxy-galaxy lensing is much less sensitive than cosmic
shear to such systematics due to the circular averaging over
lens-source pairs, as thoroughly investigated for CFHTLenS by Velander
et al.\ (2014), we do not remove these pointings from our analysis.
We explicitly checked that any difference in the galaxy-galaxy lensing
statistics between these choices was within the range of statistical
fluctuations.

The {\it lens}fit pipeline returns measured ellipticity components
$(e_1, e_2)$ for each source, together with an approximately optimal
weight $w^s$, a combination of the variance in the intrinsic
ellipticity and the measurement error due to photon noise (Miller et
al.\ 2013).  We note that the sign of $e_2$ listed in the source
catalogues must be reversed in our cross-correlation pipeline, because
the positive $x$-direction of pixel coordinates lies in the negative
R.A.\ direction.  We cut the catalogue to sources with weights $w^s >
0$ which lie in unmasked areas of data.  No magnitude cuts are
applied, although fainter galaxies are downweighted by lower values of
$w^s$.  We also do not apply a star-galaxy separation cut, since stars
are already assigned negligible weights by {\it lens}fit.  The
effective source density for lensing analyses (following Heymans et
al.\ 2012) is defined by
\begin{equation}
n_{\rm eff} = \frac{1}{A_{\rm eff}} \, \frac{ \left( \sum_i w^s_i
  \right)^2 }{\sum_i (w^s_i)^2} ,
\label{eqsourcedens}
\end{equation}
where $A_{\rm eff}$ is the effective (unmasked) area.  The values
derived for (W1, W4) are $n_{\rm eff} = (14.0, 13.1)$ arcmin$^{-2}$.
(We note that Chang et al.\ 2013 provide another possible definition
of $n_{\rm eff}$.)

The {\small BPZ} photometric-redshift pipeline returns a full redshift
probability distribution $p_{\rm BPZ}(z)$ for each source, binned in
intervals of $\Delta z = 0.05$ in the range $0 < z < 3.5$, and we make
use of this full information when computing the lensing signal.  For
purposes of binning sources in photo-$z$ slices, we use the
maximum-likelihood redshift value $z_B$ of these distributions.  The
source redshift distribution $p_s(z)$, which is required for
cosmological modelling and for building the mock catalogues below, is
constructed by stacking the $p_{\rm BPZ}(z)$ distributions for each
source, weighting by the {\it lens}fit weights,
\begin{equation}
p_s(z) \propto \sum_i w^s_i \, p_{{\rm BPZ},i}(z)
\end{equation}
We do not apply any photo-$z$ cuts in our fiducial analysis, and
demonstrate that our results are insensitive to such cuts.

We note that Miyatake et al.\ (2015) and More et al.\ (2015) recently
presented related galaxy-galaxy lensing measurements for
CFHTLenS-CMASS, with the aim of understanding the properties of the
dark matter haloes traced by the lenses, and performing joint
cosmological and halo occupation model fits.

\subsection{RCSLenS}

The 2nd Red Sequence Cluster Survey (RCS2, Gilbank et al.\ 2011) is a
nearly 800 deg$^2$ imaging survey in three optical bands $g' r' z'$
carried out with the CFHT.  The primary survey area is divided into 13
well-separated patches on the sky, each with an area ranging from 20
to 100 deg$^2$.  Nine of these fields (with properties listed in Table
\ref{tabdata}) overlap with the spectroscopic-redshift data used in
this analysis.  The main lensing analysis is performed using the $r'$
band data, with limiting magnitude $r' \approx 24.3$.  Around
two-thirds of the RCS2 area has also been imaged in the $i'$ band.

The RCS2 team have presented a series of investigations of
galaxy-galaxy lensing using these data, probing the occupation and
shapes of dark matter halos (van Uitert et al.\ 2011, 2012, Cacciato
et al.\ 2014) and the connection to galaxy luminosity, stellar mass
and velocity dispersion (van Uitert et al.\ 2013, 2015).  We focus
instead on cosmological applications of this dataset.

The RCSLenS dataset\footnote{\tt http://www.rcslens.org} is a lensing
re-analysis of the RCS2 imaging data constructed by applying the same
processing pipeline for shape measurement and photometric-redshift
estimation as developed for CFHTLenS (Hildebrandt et al.\ 2015).  The
effective RCSLenS source density derived using Equation
\ref{eqsourcedens}, $n_{\rm eff} \approx 5.5$ arcmin$^{-2}$, is lower
than the corresponding value for CFHTLenS due to the shallower
imaging, but the data cover a significantly wider area.  {\small BPZ}
photometric redshifts are only derived for pointings containing 4
optical bands $g' r' i' z'$, which correspond to about two-thirds of
the survey area (with statistics for individual fields listed in Table
\ref{tabdata}).  We note that the absence of $u^*$ band imaging in
RCSLenS causes slightly larger photo-$z$ errors and a greater outlier
fraction at low redshifts, owing to more serious colour-redshift
degeneracies.

The RCSLenS shape catalogues have been subject to a ``blinding''
process to prevent confirmation bias in analysis (Hildebrandt et
al.\ 2015, see also Kuijken et al.\ 2015 for a full description of the
same process).  In brief, the catalogues contain four sets of
ellipticity data, one of which is the true data, whilst the other
three sets have been manipulated by a trusted external colleague by
different amounts unknown to the science team, sufficient to prevent
confirmation bias but not so extreme as to render those data
suspiciously discrepant.  All scientific analyses are repeated for
each of the 4 blindings.  Once the results are ready for publication,
the blinding key is provided by the external, after which the science
team can make no further undocumented modifications to analysis
procedures.

\subsection{WiggleZ Dark Energy Survey}

The WiggleZ Dark Energy Survey (Drinkwater et al.\ 2010) is a
large-scale galaxy redshift survey of bright emission-line galaxies
over the redshift range $z < 1$ with median redshift $z \approx 0.6$,
which was carried out at the Anglo-Australian Telescope at Siding
Spring, Australia, between August 2006 and January 2011.  In total, of
order $200{,}000$ redshifts were obtained, covering of order 1000
deg$^2$ of equatorial sky divided into seven well-separated regions,
labelled as (00, 01, 03, 09, 11, 15, 22) by their location on the sky
in hours of R.A.

WiggleZ galaxies were selected for observation using colour and
magnitude cuts from a combination of optical and UV imaging.  The
optical imaging employed was from the SDSS in the Northern Galactic
Pole and from the RCS2 survey in the Southern Galactic Pole, spanning
a total of 6 RCS2 fields (0047, 0310, 1514, 2143, 2329, 2338).  The
WiggleZ dataset is therefore very well-suited for providing lenses for
cross-correlation with RCSLenS sources.  The 0047, 1514, 2143 and 2329
fields are also covered by BOSS, producing higher signal-to-noise
galaxy-galaxy lensing measurements in these cases, and therefore to
avoid the complication of the covariance we focus on analyzing the 2
remaining fields, 0310 and 2338.  When measuring the galaxy-galaxy
lensing signal, we also cut the lens and source distributions to the
subset of MegaCam pointings containing both lenses and sources (the
``overlap area'') listed in Table \ref{tabdata}.  After these cuts, we
used a total of $5{,}585$ WGZLoZ lenses and $8{,}298$ WGZHiZ lenses in
our analysis.  The WiggleZ lens selection function within each region
was determined using the methods described by Blake et al.\ (2010).
The average galaxy bias factor of the WiggleZ lenses is $b \approx 1$.

\begin{figure}
\begin{center}
\resizebox{8cm}{!}{\rotatebox{270}{\includegraphics{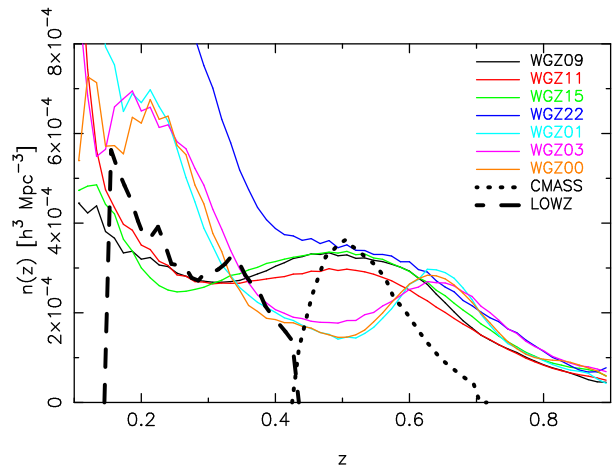}}}
\end{center}
\caption{Number density distribution with redshift of lenses in each
  WiggleZ survey region and in the BOSS LOWZ/CMASS samples.  The
  redshift distribution differs between WiggleZ regions because of
  varying colour/magnitude selection and completeness of spectroscopic
  follow-up.}
\label{figndens}
\end{figure}

Figure \ref{figndens} plots the number density distribution with
redshift of the WiggleZ lens samples in the different survey regions.
The typical number density is $n \approx 3 \times 10^{-4} \, h^3$
Mpc$^{-3}$ at the effective redshift $z \approx 0.6$, with a tail to
lower number densities at high redshift ($1 \times 10^{-4} \, h^3$
Mpc$^{-3}$ at $z \approx 0.9$) and higher number densities at low
redshift ($6 \times 10^{-4} \, h^3$ Mpc$^{-3}$ at $z \approx 0.1$).
Some variation amongst WiggleZ regions is evident, due to differences
in the colour/magnitude selection and completeness of the
spectroscopic follow-up (for full details, see Drinkwater et
al.\ 2010).

\subsection{BOSS}

BOSS is a spectroscopic follow-up survey of the SDSS III imaging
survey (Eisenstein et al.\ 2011), which has obtained redshifts for over
a million galaxies covering $10{,}000$ deg$^2$.  All observations have
been carried out at the Sloan Telescope located at Apache Point
Observatory in New Mexico.  BOSS uses colour and magnitude cuts to
select two classes of galaxies to be targetted for spectroscopy, the
``LOWZ'' sample which contains red galaxies at $z < 0.43$ and the
``CMASS'' sample which is designed to be approximately stellar-mass
limited for $z > 0.43$.  We use the data catalogues provided by the
SDSS 10th Data Release (DR10); full details of these catalogues are
given by Eisenstein et al.\ (2011), Dawson et al.\ (2013) and Anderson
et al.\ (2014).

Following the practice of the BOSS science papers we cut the LOWZ
sample to $0.15 < z < 0.43$ and the CMASS sample to $0.43 < z < 0.7$,
in order to avoid redshift overlap and create homogeneous galaxy
samples.  The redshift distributions of these samples are shown in
Figure \ref{figndens} and are comparable to the WiggleZ dataset.  The
galaxy bias factors of the LOWZ and CMASS samples are $b \approx 1.6$
and $b \approx 1.9$, respectively (Chuang et al.\ 2014, Sanchez et
al.\ 2014), hence these objects are significantly more clustered than
WiggleZ galaxies.

In order to correct for the effects of redshift failures, fibre
collisions and other known systematics affecting the angular
completeness (correlated with the stellar density and seeing), BOSS
galaxies are assigned completeness weights (as specified in Equation
18 of Anderson et al.\ 2014).  We included these weights in our
determination of the auto-correlation function of the galaxies, but
they do not affect the galaxy-galaxy lensing measurements because they
are uncorrelated with the background shapes.

\begin{figure*}
\begin{center}
\resizebox{12cm}{!}{\rotatebox{0}{\includegraphics{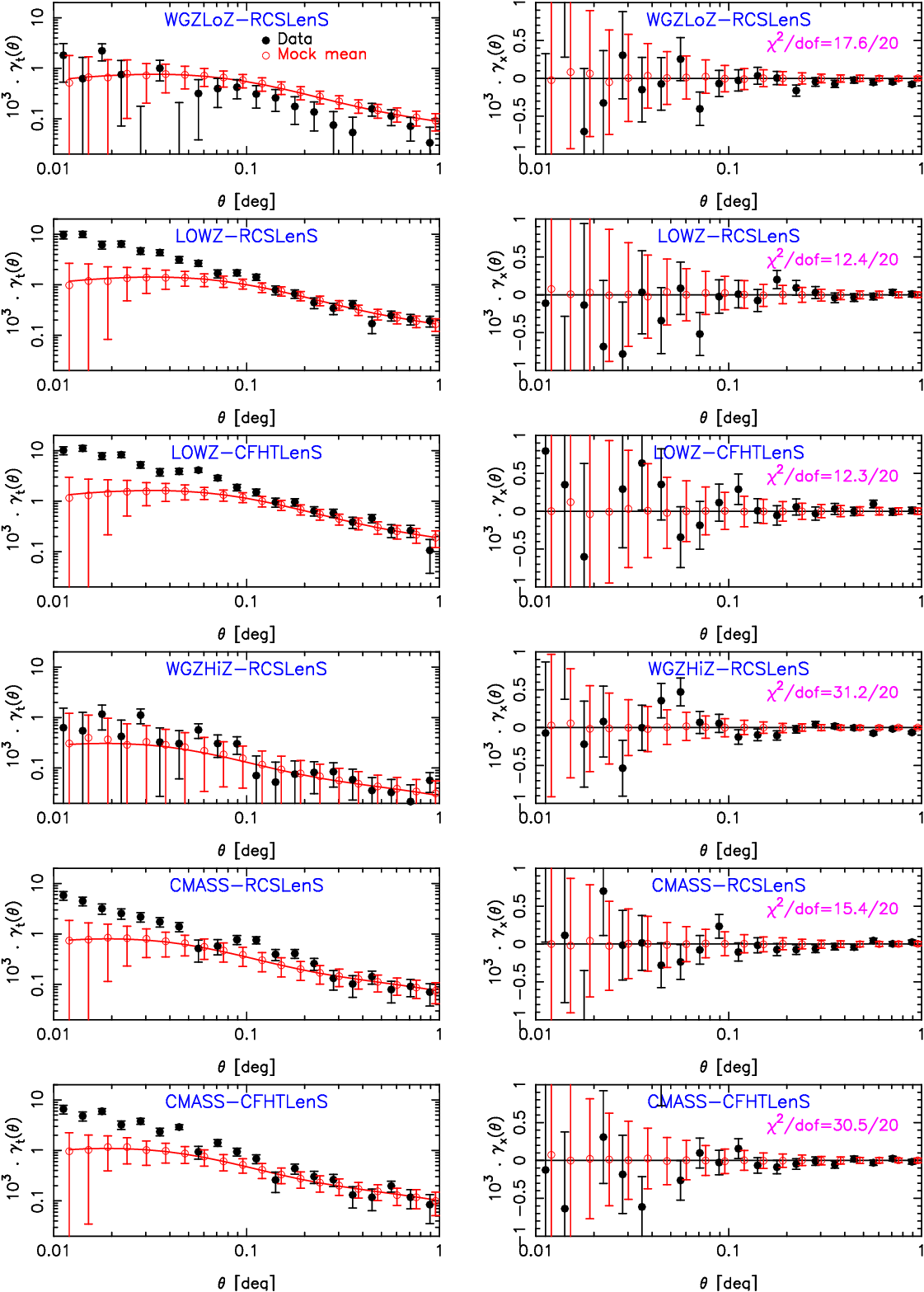}}}
\end{center}
\caption{Measurements of $\gamma_t(\theta)$ (left column) and
  $\gamma_\times(\theta)$ (right column) for the cross-correlation of
  different combinations of source-lens datasets.  We show results for
  both the data (black solid circles) and mock mean (red open
  circles), with the errors based on measurements for a set of 374
  mock catalogues.  The overplotted model is the expectation for the
  mock mean based on the input cosmology of the simulations.  $\chi^2$
  statistics are quoted for the $\gamma_\times$ measurements with
  respect to a model of zero, with number of degrees of freedom ${\rm
    dof}=20$.}
\label{figgtgx}
\end{figure*}

As noted in Table \ref{tabdata}, the BOSS sample overlaps 2 CFHTLenS
fields (W1 and W4) and 7 RCSLenS fields (0047, 0133, 1303, 1514, 1645,
2143, 2329).  We utilized a total of $12{,}172$ LOWZ lenses and
$24{,}135$ CMASS lenses in our analysis.

\section{Galaxy-galaxy lensing measurements}
\label{secmeas}

In this Section we describe our estimation of the stacked tangential
shear around the lenses as a function of angular separation and the
differential surface density around the lenses as a function of the
projected physical separation.  In the latter case we use the full
source photometric-redshift probability distribution when computing
the lensing signal.

\subsection{Average tangential shear $\gamma_t(\theta)$}
\label{secgtest}

The ellipticity components $(e_{1,i}, e_{2,i})$ of source $i$ relative
to the positive $x$-axis are defined by
\begin{equation}
(e_1, e_2) = \left( \frac{r-1}{r+1} \right) (\cos{2\psi}, \sin{2\psi})
  ,
\end{equation}
where $\psi$ is the position angle of a galaxy with axial ratio $r$
measured anti-clockwise from the positive $x$-axis.  These ellipticity
components can be rotated to new values $[e_t(i,j), e_\times(i,j)]$
relative to a line connecting source $i$ and lens $j$ by
\begin{equation}
e_t(i,j) = - e_{1,i} \cos{2\phi(i,j)} - e_{2,i} \sin{2\phi(i,j)}
\end{equation}
and
\begin{equation}
e_\times(i,j) = e_{1,i} \sin{2\phi(i,j)} - e_{2,i} \cos{2\phi(i,j)} ,
\end{equation}
where $\phi(i,j)$ is the angle of the line connecting source $i$ and
lens $j$ to the positive $x$-axis (in the range $-90^\circ < \phi <
90^\circ$).  Our galaxy-galaxy lensing estimators for the tangential
shear $\gamma_t$ and cross shear $\gamma_\times$ are then:
\begin{equation}
\gamma_t(\theta) = \frac{\sum_{{\rm sources} \, i} \sum_{{\rm lenses}
    \, j} w^s_i \, w^l_j \, e_t(i,j) \, \Theta(i,j)}{\sum_{{\rm
      sources} \, i} \sum_ {{\rm lenses} \, j} w^s_i \, w^l_j \,
  \Theta(i,j)}
\end{equation}
and
\begin{equation}
\gamma_\times(\theta) = \frac{\sum_{{\rm sources} \, i} \sum_{{\rm
      lenses} \, j} w^s_i \, w^l_j \, e_\times(i,j) \,
  \Theta(i,j)}{\sum_{{\rm sources} \, i} \sum_{{\rm lenses} \, j}
  w^s_i \, w^l_j \, \Theta(i,j)} ,
\end{equation}
where $w^s_i$ is the {\it lens}fit weight of source $i$, $w^l_j$ is
the weight of lens galaxy $j$ which we set to the completeness weights
for BOSS galaxies and $w^l = 1$ for WiggleZ galaxies (since in that
case the completeness is incorporated in the random catalogues),
$\Theta(i,j)$ is equal to 1 if the angular separation between source
$i$ and lens $j$ lies in bin $\theta$ and equal to 0 otherwise, and
the sums are taken over unique pairs.  We performed measurements in 20
logarithmically-spaced angular bins from $\theta = 0.01^\circ$ to
$1^\circ$.  Multiplicative shear bias corrections are computed and
applied as discussed in Section \ref{seccal}.  The cross shear
$\gamma_\times$ should be consistent with zero signal for weak
gravitational lensing but potentially non-zero for distortions due to
systematic errors.

Figure \ref{figgtgx} shows the resulting measurements of the
tangential and cross shear for the six different combinations of
source-lens datasets.  The measurements are performed in the
individual fields listed in Table \ref{tabdata} and then combined
together using inverse-variance weighting.  The errors in the
measurements are obtained by applying the same analysis pipeline to a
set of 374 mock catalogues, which are constructed by the process
described in Section \ref{secsim}, and appropriately scaling the
standard deviation of the mock statistics.  In Figure \ref{figgtgx} we
plot measurements for both the data (black solid circles) and the mean
of the mock catalogues (red open circles), where the solid line is the
model prediction for the mocks based on the input cosmology of the
simulations.  In the absence of systematic effects, the cross shear
should be consistent with zero.  Values of the chi-squared statistic
are shown for each source-lens combination, determined for a model
$\gamma_\times = 0$ using the covariance matrix built from the mock
catalogues.  Our measurements are consistent with this expectation.

At small scales, for values of $\theta$ corresponding to projected
physical separations $R < 2 \, h^{-1}$ Mpc, the amplitude of
galaxy-galaxy lensing in the mock falls below that measured in the
data.  This arises because of halo occupation effects on small scales
particularly affecting Luminous Red Galaxies; it presents no concern
for our analysis, in which physical effects originating at small
scales $R < R_0$ are explicitly suppressed.

\subsection{Differential surface density $\Delta \Sigma(R)$}
\label{secdelsigest}

In Section \ref{secdelsigtheory} we demonstrated that for a relatively
narrow lens distribution, $\Delta \Sigma(R, z_l)$ could be estimated
from the tangential galaxy shear via Equation \ref{eqgtdelsig3}, which
we re-write here as
\begin{equation}
\Delta \Sigma(R, z_l) = \gamma_t(\theta) / \left(
\overline{\Sigma_c^{-1}} \right) ,
\label{eqgtdelsig4}
\end{equation}
where
\begin{eqnarray}
\overline{\Sigma_c^{-1}} &\equiv& \int_{z_l}^\infty dz_s \, p_s(z_s)
\, \Sigma_c^{-1}(z_l,z_s) \nonumber \\ &=& \frac{4 \pi G \, (1 + z_l)
  \, \chi(z_l)}{c^2} \int_{z_l}^\infty dz_s \, p_s(z_s) \left[ 1 -
  \frac{\chi(z_l)}{\chi(z_s)} \right] .
\label{eqinvsigc}
\end{eqnarray}
We implemented this calculation for each lens-source pair by using the
source photometric-redshift probability distribution, $p_s(z) = p_{\rm
  BPZ}(z)$.  Therefore we estimate $\Delta \Sigma$ using only that
lensing data for which photometric-redshift information is available,
as listed in Table \ref{tabdata}.

\begin{figure*}
\begin{center}
\resizebox{13cm}{!}{\rotatebox{270}{\includegraphics{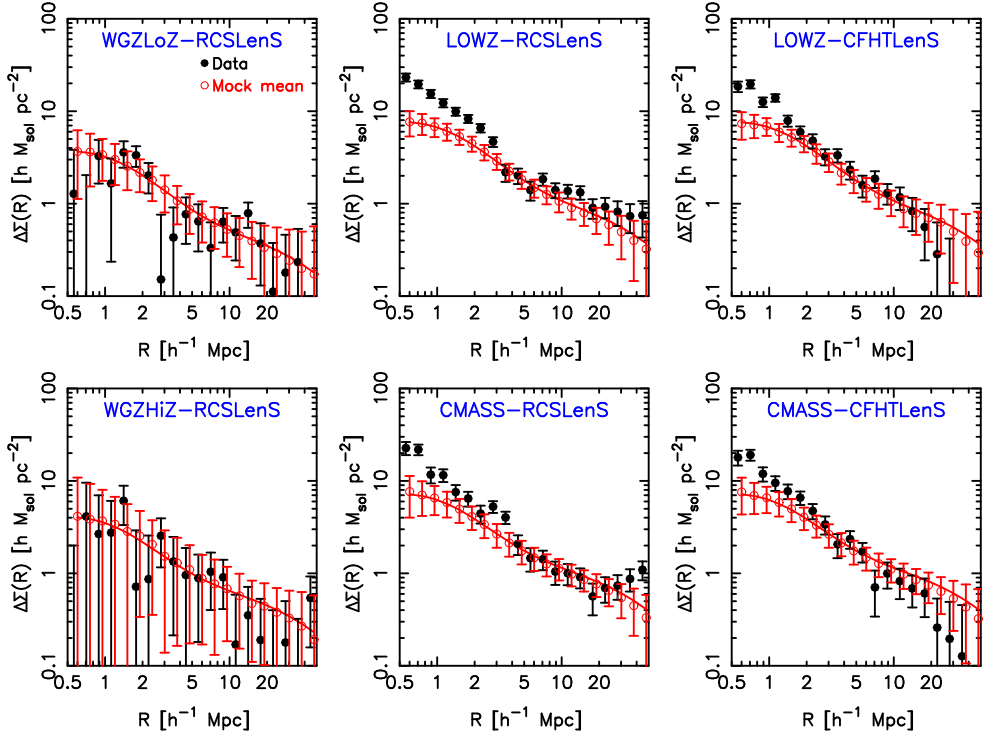}}}
\end{center}
\caption{$\Delta \Sigma(R)$ measured for the cross-correlation of
  different combinations of source-lens datasets.  We show results for
  the data and the mock mean.  The overplotted model is included only
  for comparison with the mock mean.}
\label{figdelsig}
\end{figure*}

\begin{figure*}
\begin{center}
\resizebox{13cm}{!}{\rotatebox{270}{\includegraphics{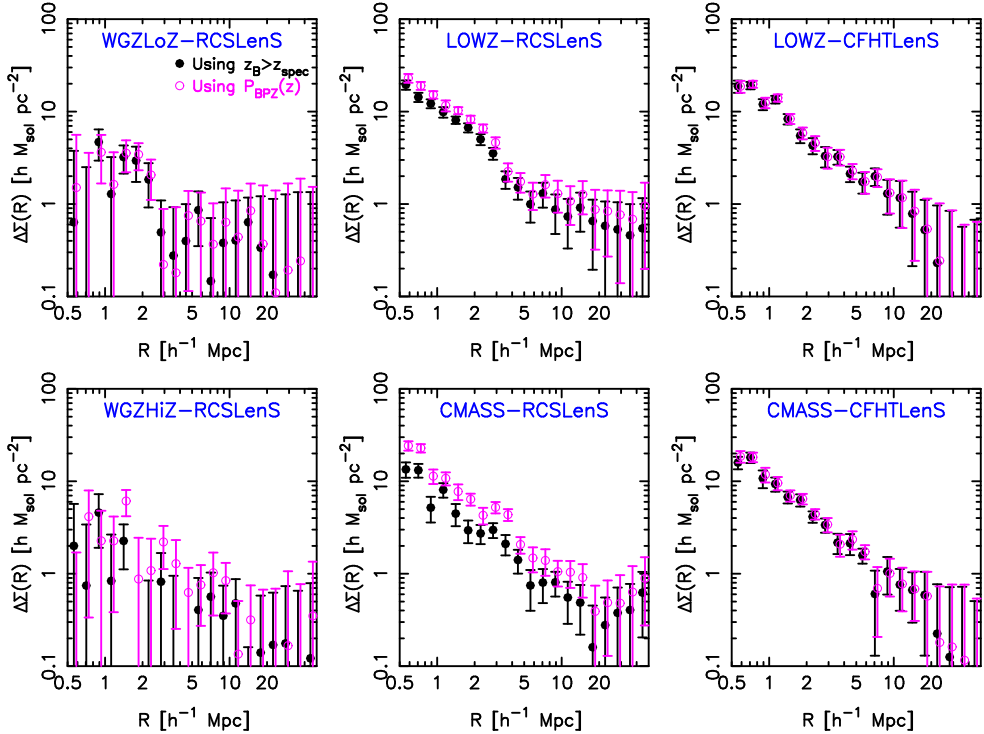}}}
\end{center}
\caption{A comparison of the measurements of $\Delta \Sigma(R)$ for
  different combinations of source-lens datasets using the approximate
  estimator of Equation \ref{eqdelsigest2} based only on the best
  source photometric redshift $z_B$, plotted as the solid (black)
  circles, and the unbiased estimator of Equation \ref{eqdelsigest1},
  based on the full redshift probability distribution $p_{\rm BPZ}(z)$
  of each source, plotted as the open (magenta) circles.  For the
  purposes of this plot the errors were determined by jack-knife
  re-sampling.}
\label{figpzest}
\end{figure*}

For our determinations of $\overline{\Sigma_c^{-1}}$ using Equation
\ref{eqinvsigc}, and conversions of source-lens angular separations
$\theta$ to projected separations $R = \theta \, \chi(z_l)$, we
adopted a fiducial flat $\Lambda$CDM cosmological model with matter
density $\Omega_m = 0.27$.  Our motivation for this choice, which is
in better agreement with the fits to the Cosmic Microwave Background
fluctuations using WMAP satellite data (Komatsu et al.\ 2011) rather
than the later {\it Planck} satellite data ({\it Planck} collaboration
2015b), is to ensure consistency with the fiducial cosmological model
adopted for the RSD analyses of the WiggleZ and BOSS data (Blake et
al.\ 2011, Sanchez et al.\ 2014), which would be subject to
Alcock-Paczynski distortion in different models.  Adopting the higher
value of $\Omega_m$ preferred by the {\it Planck} analysis would not
produce a significant change in our measurements compared to the
statistical errors.

When averaging the estimates of Equation \ref{eqgtdelsig4} over the
lens-source pairs, each source must now be inverse-variance weighted
as $w^s \left( \overline{\Sigma_c^{-1}} \right)^2$ and hence
our final estimator is
\begin{equation}
\Delta \Sigma(R) = \frac{\sum_{{\rm sources} \, i} \sum_{{\rm lenses}
    \, j} w^s_i \, w^l_j \, \overline{\Sigma_c^{-1}}_{ij} \, e_t(i,j)
  \, \Theta(i,j)}{\sum_{{\rm sources} \, i} \sum_ {{\rm lenses} \, j}
  w^s_i \, w^l_j \, \left[ \overline{\Sigma_c^{-1}}_{ij} \right]^2 \,
  \Theta(i,j)}
\label{eqdelsigest1}
\end{equation}
(see also Miyatake et al.\ 2015).  This estimator corrects for the
dilution in lensing signal caused by the non-zero probability that a
source is situated in front of the lens, $z_s < z_l$.

We note that an approximate version of this estimator has been used in
previous studies where only the maximum-likelihood source redshift
$z_s$ is available, rather than the full probability distribution
$p_s(z_s)$.  In this approximation the lens-source pairs are
restricted to those cases for which $z_s > z_l$ and the differential
surface density for each pair is estimated as
\begin{equation}
\Delta \Sigma(R, z_l) \approx \gamma_t(\theta) \, \Sigma_c(z_l, z_s) .
\end{equation}
The inverse-variance weight of each source-lens pair is now $w^s \,
\Sigma_c^{-2}$ such that the approximate estimator reads
\begin{equation}
\Delta \Sigma(R) = \frac{\sum_{{\rm sources} \, i} \sum_{{\rm lenses}
    \, j} w^s_i \, w^l_j \, \Sigma_{c,ij}^{-1} \, e_t(i,j) \,
  \Theta(i,j)}{\sum_{{\rm sources} \, i} \sum_ {{\rm lenses} \, j}
  w^s_i \, w^l_j \, \Sigma_{c,ij}^{-2} \, \Theta(i,j)} .
\label{eqdelsigest2}
\end{equation}
This relation will contain a bias as far as $\Sigma_c \ne
1/\overline{\Sigma_c^{-1}}$, although that bias can be corrected
through comparison with spectroscopic catalogues (Nakajima et
al.\ 2012).

Figure \ref{figdelsig} shows the measurement of $\Delta \Sigma(R)$
using Equation \ref{eqdelsigest1} in 20 logarithmically-spaced bins in
$R$ from $0.5$ to $50 \, h^{-1}$ Mpc, for the six different
combinations of source-lens datasets, displaying results for both the
data and the mock mean.  As above, the errors in these measurements
are obtained using the 374 mock catalogues introduced in Section
\ref{secsim}.  Our choice of this binning scheme allows us to search
for scale-dependence in the measurement whilst retaining good
signal-to-noise in each bin.

\begin{figure}
\begin{center}
\resizebox{8cm}{!}{\rotatebox{270}{\includegraphics{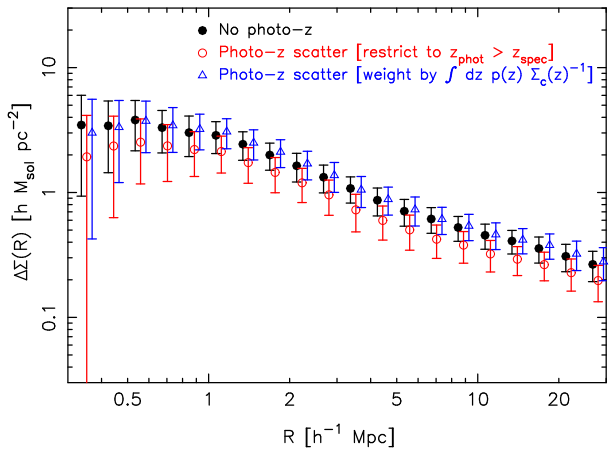}}}
\end{center}
\caption{Verification that the estimator of Equation
  \ref{eqdelsigest1} successfully recovers the input value of $\Delta
  \Sigma(R)$ in mock catalogues.  A toy photo-$z$ model is applied to
  the mocks, in which redshifts are scattered in accordance with a
  Gaussian distribution of zero mean and standard deviation $0.2
  \times (1+z)$.  The differential surface density is then estimated
  using both Equation \ref{eqdelsigest2}, restricting the calculation
  to source-lens pairs for which $z_{\rm phot} > z_{\rm spec}$ (red
  open circles), and Equation \ref{eqdelsigest1}, which uses the full
  information of the photo-$z$ probability distribution (green open
  triangles).  These determinations are compared to the measurements
  without application of photometric redshifts (black solid circles).
  In all cases the mock mean is plotted, where the error represents
  the standard deviation of the mocks.}
\label{figpzesttest}
\end{figure}

We now consider in more detail the difference in results produced by
the unbiased estimator of Equation \ref{eqdelsigest1} and the
approximate estimator of Equation \ref{eqdelsigest2}.  First, Figure
\ref{figpzest} compares measurements of $\Delta \Sigma(R)$ using the
two estimators.  We conclude that the approximate estimator contains a
significant systematic bias, especially when using photometric
redshifts with precision characteristic of the RCSLenS data.
Secondly, Figure \ref{figpzesttest} presents a verification of our
successful recovery of $\Delta \Sigma(R)$ in the presence of
photometric-redshift errors, using the mock catalogues described in
Section \ref{secsim}.  A toy photo-$z$ model is applied to the mock
catalogues, in which redshifts are scattered in accordance with a
Gaussian distribution of zero mean and standard deviation $0.2 \times
(1+z)$, and the differential surface density is then estimated using
both Equation \ref{eqdelsigest1} and Equation \ref{eqdelsigest2}.  The
measurements in which the full source redshift probability
distributions are used agree well with an analysis in which the
photo-$z$ scatter is not applied.

\begin{figure}
\begin{center}
\resizebox{8cm}{!}{\rotatebox{270}{\includegraphics{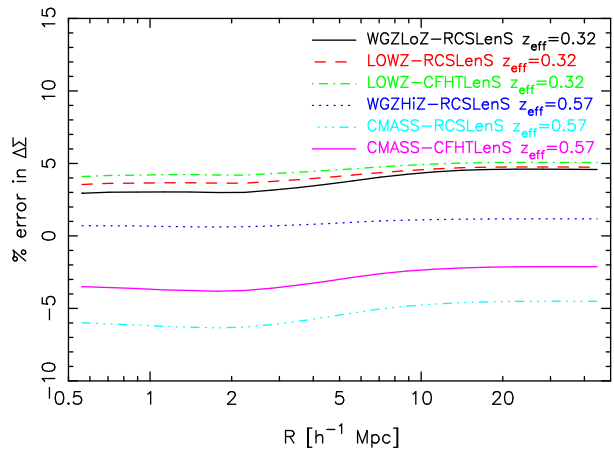}}}
\end{center}
\caption{The systematic error in determination of $\Delta \Sigma$ that
  results in the application of the narrow-lens approximation of
  Equation \ref{eqgtdelsig3} at the effective redshift of the RSD
  measurements of each lens sample.}
\label{fignarrowlens}
\end{figure}

Finally, Figure \ref{fignarrowlens} assesses how accurately the narrow
lens approximation described in Section \ref{secdelsigtheory} holds
for the specific lens and source redshift distributions of our
samples.  The amplitude of the systematic error is calculated by
comparing the value of $\Delta \Sigma(R)$ that would be inferred at
the effective redshifts at which RSD is measured for the different
lens samples by substituting Equation \ref{eqgtdelsig2} in Equation
\ref{eqgtdelsig3}, with the fiducial value of $\Delta \Sigma(R)$
evaluated using Equation \ref{eqxigm}.  We find that the resulting
amplitude of the systematic error in $\Delta \Sigma$ (and consequently
$E_G$) is about $5\%$, which is significantly smaller than the
statistical error in $E_G$ of around $20\%$.  We conclude that the
narrow-lens approximation is acceptable for our analysis.

We note (without further calculation) an additional issue for
estimating $E_G$ that could occur for wide lens redshift slices.  The
weighting applied to each source-lens pair when estimating $\Delta
\Sigma(R)$, $\overline{\Sigma_c^{-1}}$, contains a dependence on the
lens redshift which may induce a systematic difference in the relative
weighting of lens galaxies to that used in the measurement of their
projected auto-correlation function $w_p(R)$, which would become
important if the bias of the lenses evolved across the redshift slice.
However, the bias of both WiggleZ and BOSS lenses is a fairly flat
function of redshift across these ranges (Blake et al.\ 2009, Anderson
et al.\ 2014).

\subsection{Calibration and random-catalogue corrections}
\label{seccal}

\begin{figure*}
\begin{center}
\resizebox{14cm}{!}{\rotatebox{270}{\includegraphics{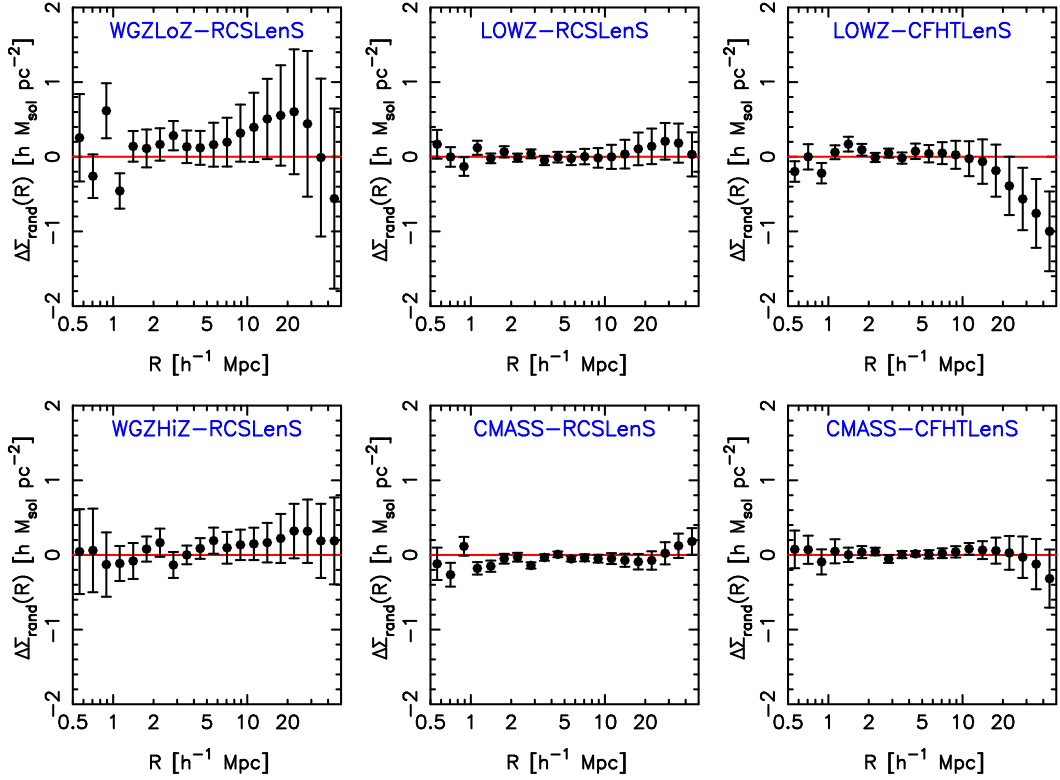}}}
\end{center}
\caption{The cross-correlation between shapes and random lenses,
  $\Delta \Sigma_{\rm rand}(R)$, measured for the different
  combinations of source and lens samples.  The results are averaged
  over 40 random catalogues.  A jack-knife error is plotted, noting
  that this error is only accurate for small scales and is a
  significant over-estimate for the largest scales.  This systematic
  correction is subtracted from the data, which is standard practice
  for calibrating galaxy-galaxy lensing measurements (see for example
  Mandelbaum et al.\ 2005).}
\label{figsigrand}
\end{figure*}

Bias in ellipticity measurements may be described by a linear
combination of a multiplicative error $m$ and an additive error $c$
such that
\begin{equation}
e_i^{\rm obs} = (1+m_i) \, e_i^{\rm true} + c_i + {\rm Noise} ,
\label{eqshearbias}
\end{equation}
where $i=1,2$ denotes the two ellipticity components.

Velander et al.\ (2014) show that the additive shear bias has a
negligible effect for galaxy-galaxy lensing measurements with CFHTLenS
data.  We follow the standard practice of performing this calibration
by subtracting the correlation with a random lens catalogue as
described below, which represents an empirical calculation of any
additive shear residual.

The multiplicative shear bias correction $m$ defined in Equation
\ref{eqshearbias} has been modelled for {\it lens}fit processing of
MegaCam data using a shear recovery test based on galaxy image
simulations as a function of source signal-to-noise $SN$ and size in
pixels $r$:
\begin{equation}
m(SN,r) = \frac{\beta}{{\rm log}_{10}(SN)} \exp{(- r \, SN \, \alpha)} ,
\label{eqmult}
\end{equation}
where $\alpha = 0.057$ and $\beta = -0.37$ (Miller et al.\ 2013).  The
same correction applies to both $e_1$ and $e_2$, and is greater for
sources with smaller signal-to-noise ratios and sizes.  We propagated
this correction into the measurement of $\gamma_t(\theta)$ by
evaluating
\begin{equation}
K(\theta) = \frac{\sum_{{\rm sources} \, i} \sum_{{\rm lenses} \, j}
  w_i^s \, w_j^l \, m_i \, \Theta(i,j)}{\sum_{{\rm sources} \, i}
  \sum_{{\rm lenses} \, j} w_i^s \, w_j^l \, \Theta(i,j)} ,
\label{eqktheta}
\end{equation}
such that the corrected measurement is given by
\begin{equation}
\gamma_t^{\rm corrected}(\theta) = \frac{\gamma_t^{\rm
    uncorrected}(\theta)}{1 + K(\theta)} .
\end{equation}
We note that this multiplicative shear bias correction must be applied
in a global fashion, not on an individual source basis, due to its
correlation with the measured source ellipticity values owing to its
dependence on $SN$ and $s$ in Equation \ref{eqmult}.

The analogous formula for the correction to be applied to $\Delta
\Sigma(R)$ is
\begin{equation}
K(R) = \frac{\sum_{{\rm sources} \, i} \sum_{{\rm lenses} \, j} w^s_i
  \, w^l_j \, m_i \, \Sigma_{c,ij}^{-1} \, \Theta(i,j)}{\sum_{{\rm
      sources} \, i} \sum_ {{\rm lenses} \, j} w^s_i \, w^l_j \,
  \Sigma_{c,ij}^{-2} \, \Theta(i,j)} .
\label{eqkr}
\end{equation}
Equation \ref{eqkr} assumes that we are using the approximate
estimator of Equation \ref{eqdelsigest2}; when instead employing the
unbiased estimator of Equation \ref{eqdelsigest1} we should replace
$\Sigma_{c,ij}^{-1}$ by $\overline{\Sigma_c^{-1}}_{ij}$.  Figure
\ref{figcorrmult} in Appendix \ref{secsyscal} displays the corrections
measured for the different source and lens samples; we find that $K
\approx -0.06$ independent of scale, such that the amplitude of the
galaxy-galaxy lensing signal must be boosted by $\approx 6\%$.

The shear estimated by these methods might still be contaminated by
two further effects: (1) large-scale residual shape-measurement
systematics, such as an imperfectly-modelled optical distortion across
the camera field; (2) the physical association of source and lens
galaxies, for example sharing the same dark matter halo, diluting the
cross-correlation signal since these sources will not be lensed and
yet may have been scattered to higher redshifts by photo-$z$ error.
The importance of both of these effects can be determined using random
lenses sampled from the same selection function as the data, where we
averaged over 40 random catalogues in these analyses (checking that
our final results were insensitive to the number of catalogues and the
uncertainty in the correction).

First, we re-ran the shear measurement replacing the data lenses with
the random lens catalogues.  The result for the differential projected
surface density, which we denote $\Delta \Sigma_{\rm rand}(R)$, is
displayed in Figure \ref{figsigrand} as the combined signals for the
different combinations of source and lens samples.  In the absence of
residual coherent tangential distortions the signal should be
consistent with zero; however, a significant correction is obtained at
large scales for some of our samples, in particular CFHTLenS, and is
always subtracted from our measurements of $\Delta \Sigma(R)$.  We
investigate the effect of physically-associated sources in Appendix
\ref{secboost}, concluding that it is not significant for the scales
we study.  We compared our results to those presented by Miyatake et
al.\ (2015).  These authors find a more significant large-scale signal
in $\Delta \Sigma_{\rm rand}(R)$ for CMASS-CFHTLenS than we do,
although we note that they analyze a different geometry of CMASS
lenses.

\subsection{Summary of systematics tests}
\label{secsys}

We supported our science measurements with a series of tests for
potential systematic errors.  These results are described in detail in
the Appendices, and we provide a brief summary in this Section.

In Appendix \ref{secsysshear} we present a series of systematics tests
based on re-measuring $\gamma_t(\theta)$ following various
manipulations of the shear catalogue:
\begin{itemize}
\item Rotation of the sources by $45^\circ$ (i.e., $e_{1,{\rm new}} =
  e_{2,{\rm old}}$ and $e_{2,{\rm new}} = - e_{1,{\rm old}}$).
\item Randomizing the shear values amongst the source catalogue.
\item Replacing the lens catalogue by a random catalogue.
\end{itemize}
The results of these tests are consistent with $\gamma_t = 0$, as
expected.

Systematic errors in the {\small BPZ} photometric redshift
distributions would imprint errors in the determination of $\Delta
\Sigma(R)$ (given that these full probability distributions are used
in the measurement, as described above).  In order to search for such
effects we performed a ``scaling test'' of the galaxy-galaxy lensing
signal measured for the same set of foreground lenses using background
sources in a series of different photo-$z$ slices spanning the range
$0 < z_B < 1.6$.  The results are detailed in Appendix
\ref{secsysphotz}.  We find that a consistent lens singular isothermal
sphere velocity dispersion is obtained for all source photo-$z$ slices
with the possible exception of $0 < z_B < 0.2$, which comprises a
negligible number of galaxies.  As a final test, we repeated the $E_G$
measurement using sources in different $z_B$ ranges; these results are
discussed in Section \ref{seceg} and demonstrate that the measurement
is robust to these choices.

Another potential source of systematic error is intrinsic alignment of
the source population with respect to the foreground lenses (Blazek et
al.\ 2012), which is expected to preferentially diminish the average
tangential shear of red source galaxies.  In Appendix
\ref{secsysalign} we carry out shear measurements for red and blue
sub-samples to verify that the amplitude of these effects are below
the level that would impact our results; we will demonstrate this
explicitly in Section \ref{seceg} by repeating the $E_G$ measurement
for a blue sub-sample.

In addition to the systematic tests carried out in the paper, we refer
the reader to Kuijken et al.\ (2015) who apply the same data analysis
software used in this analysis to data from the Kilo Degree Survey.

\section{Simulations}
\label{secsim}

\begin{figure}
\begin{center}
\resizebox{8cm}{!}{\rotatebox{270}{\includegraphics{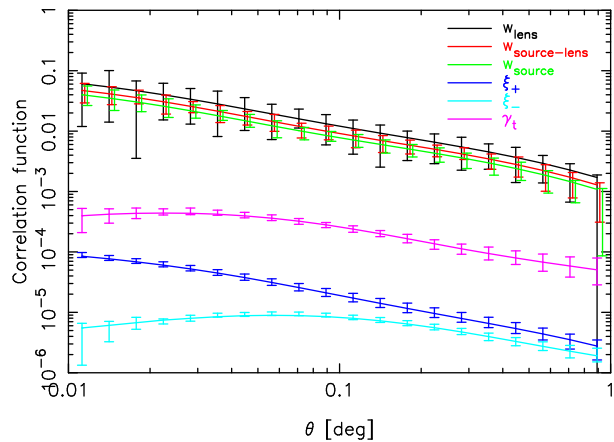}}}
\end{center}
\caption{Comparison of the mean of various clustering and lensing
  statistics measured for the mock catalogues, with the expectations
  based on the input cosmological model of the simulation and the
  source and lens redshift distributions and bias factors.  Results
  are shown for the lens and source auto-correlation functions $w_{\rm
    lens}$ and $w_{\rm source}$, the cross-correlation function
  $w_{\rm source-lens}$, the cosmic shear statistics $\xi_+$ and
  $\xi_-$, and the galaxy-galaxy lensing signal $\gamma_t$.  The error
  bars correspond to the standard deviation across the mock
  catalogues (which do not contain shape noise).}
\label{figmockmean}
\end{figure}

\subsection{Mock catalogues}

In order to test our analysis pipeline and estimate the covariance
matrix of the measurements, we created mock catalogues of each source
and lens field.  These catalogues were built from a set of 374 N-body
simulations created using methods similar to those described by
Harnois-Deraps, Vafaei \& van Waerbeke (2012).  In brief, the N-body
simulations are produced by the {\small CUBEP$^3$M} code with a
transfer function obtained from {\small CAMB} (Lewis, Challinor \&
Lasenby \ 2000) using the following cosmological parameter set: matter
density $\Omega_m = 0.2905$, baryon density $\Omega_b = 0.0473$,
Hubble parameter $h = 0.6898$, spectral index $n_s = 0.969$ and
normalization $\sigma_8 = 0.826$.  The box-size of the simulations is
$L = 505 \, h^{-1}$ Mpc; this is significantly larger than the
simulation set used for modelling the earlier CFHTLenS measurements
[$L = (147, 231) \, h^{-1}$ Mpc] such that the new simulation set is
much less affected by suppression of the large-scale variance by
finite box size.  For each simulation the density field is output at
18 snapshots, and a survey cone spanning 60 deg$^2$ is constructed by
pasting together these snapshots, where the division between adjacent
snapshots is taken as the mid-point of the cosmic distances
corresponding to the output redshifts.  The two-component shear fields
are also computed at each output redshift, by ray-tracing through the
survey cone using Born's approximation.  More details are provided by
Harnois-Deraps \& van Waerbeke (2015).

For every different survey region and lens-source survey pairing, we
converted each simulation line-of-sight into a mock catalogue by
generating:
\begin{itemize}
\item A source distribution with the surface density values given in
  Table \ref{tabdata} (adjusted for each region), using the RCSLenS or
  CFHTLenS source redshift distribution as appropriate, and
  Monte-Carlo sampling sources from the density field with bias
  $b_{\rm source} = 1$.  The source shear components are assigned by
  linearly interpolating the shear fields at the source redshift from
  the enclosing lens planes.
\item A lens distribution with the numbers given in Table
  \ref{tabdata} (adjusted for each region), using the average WiggleZ,
  LOWZ or CMASS redshift distribution as appropriate, sampled from the
  density fields with bias factors $b_{\rm lens} = (1.0, 1.6, 1.9)$
  for WiggleZ, LOWZ and CMASS, matching the bias measurements of Blake
  et al.\ 2010, Chuang et al.\ 2014 and Sanchez et al.\ 2014,
  respectively.
\end{itemize}
Sources and lenses are produced with a continuous distribution in
redshift by interpolating across the finite redshift width of each
simulation snapshot; at this stage the angular selection function is
uniform across the 60 deg$^2$ cone.  For a given bias factor $b$, the
galaxy density field $\rho_g$ is related to the mass density field
$\rho_m$ as
\begin{equation}
\frac{\rho_g}{\langle \rho_g \rangle} = 1 + b \left[
\frac{\rho_m}{\langle \rho_m \rangle} - 1 \right] .
\label{eqbg}
\end{equation}
We note that a bias model with $b > 1$ cannot be applied
self-consistently on all scales because then $\rho_g$ contains
negative regions.  We avoid these regions by imposing the condition
$\rho_g = {\rm max}(\rho_g,0)$, which reduces the effective value of
the large-scale bias by a few per cent (we settled on this approach
after experimenting with various approaches for ameliorating this
effect such as smoothing the density field before applying the
sub-sampling).

Figure \ref{figmockmean} tests the resulting mock catalogues by
measuring the average tangential shear $\gamma_t$, cosmic shear signal
$(\xi_+, \xi_-)$ and source-lens auto- and cross- angular correlation
functions $w(\theta)$ for the RCSLenS-WiggleZ mock catalogues (prior
to the application of shape noise).  The lines are the model
predictions assuming the input cosmological parameters of the
simulation, and source and lens redshift distributions and bias
factors.  The matter power spectrum is generated using the {\small
  CAMB} transfer function with non-linear contribution from {\small
  HALOFIT} (Smith et al.\ 2003).  The mock mean closely follows the
model predictions in each case.

Shape noise is applied to the source catalogues using the following
method:
\begin{itemize}
\item A complex noise $n = n_1 + n_2 \, i$ is formed for each source,
  where $n_1$ and $n_2$ are drawn from Gaussian distributions with
  standard deviation $\sigma_e$.
\item A complex shear $\gamma = \gamma_1 + \gamma_2 \, i$ is formed
  from the shear components $(\gamma_1, \gamma_2)$ obtained from the
  ray-traced shear field at the source position, linearly
  interpolating between the values at adjacent snapshot redshifts.
\item A complex noisy shear is then determined as $e = (\gamma + n)/(1
  + n \, \gamma^*)$ (Seitz \& Schneider 1997).  The components of the
  observed shear $(e_1, e_2)$ are then found as $e = e_1 + e_2 \, i$.
\end{itemize}
We used the value $\sigma_e = 0.29$ for our mocks, which is
representative of the two imaging surveys (Heymans et al.\ 2013,
Hildebrandt et al.\ 2015) albeit slightly on the conservative side.
The result of this procedure is 374 mock catalogues, matching the
global properties of the source and lens samples with a uniform
angular selection function across 60 deg$^2$.

\begin{figure*}
\begin{center}
\resizebox{13cm}{!}{\rotatebox{270}{\includegraphics{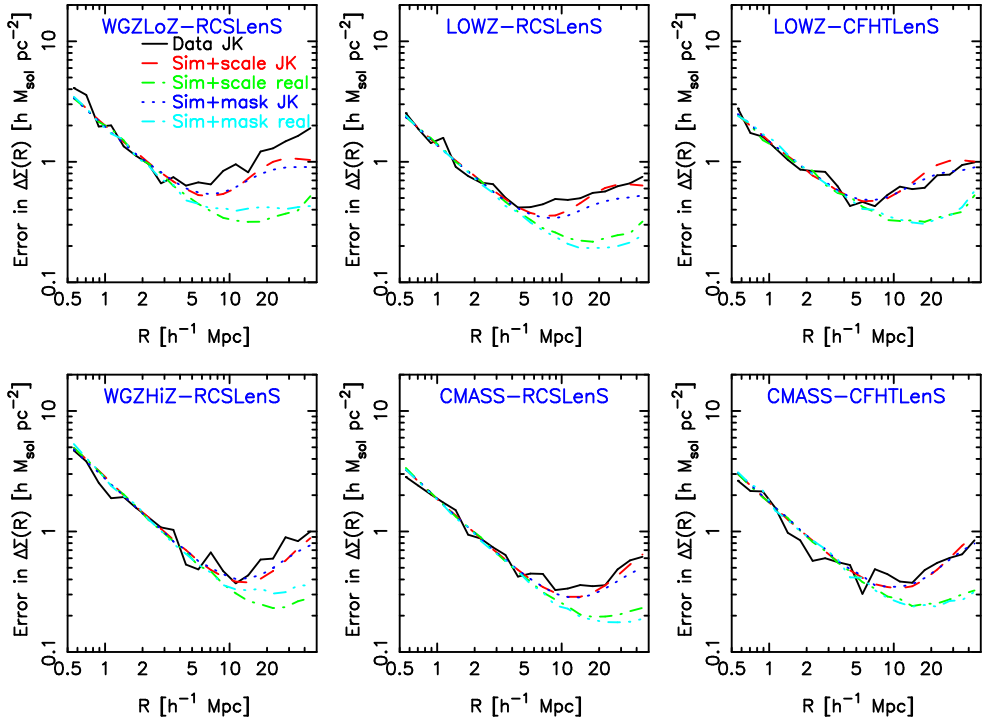}}}
\end{center}
\caption{Comparison of the errors in $\Delta \Sigma(R)$ determined by
  jack-knife re-sampling of the data (labelled Data JK), by using the
  374 simulation lines-of-sight and scaling by an effective area
  factor (comparing jack-knife re-sampling of the simulations labelled
  Sim+scale JK, and the scatter between the realizations labelled
  Sim+scale real), and by generating 149 simulations of each region
  including the full selection functions (comparing jack-knife
  re-sampling labelled Sim+mask JK, and the realization scatter
  labelled Sim+mask real).}
\label{figerrcomp}
\end{figure*}

\begin{figure*}
\begin{center}
\resizebox{13cm}{!}{\rotatebox{270}{\includegraphics{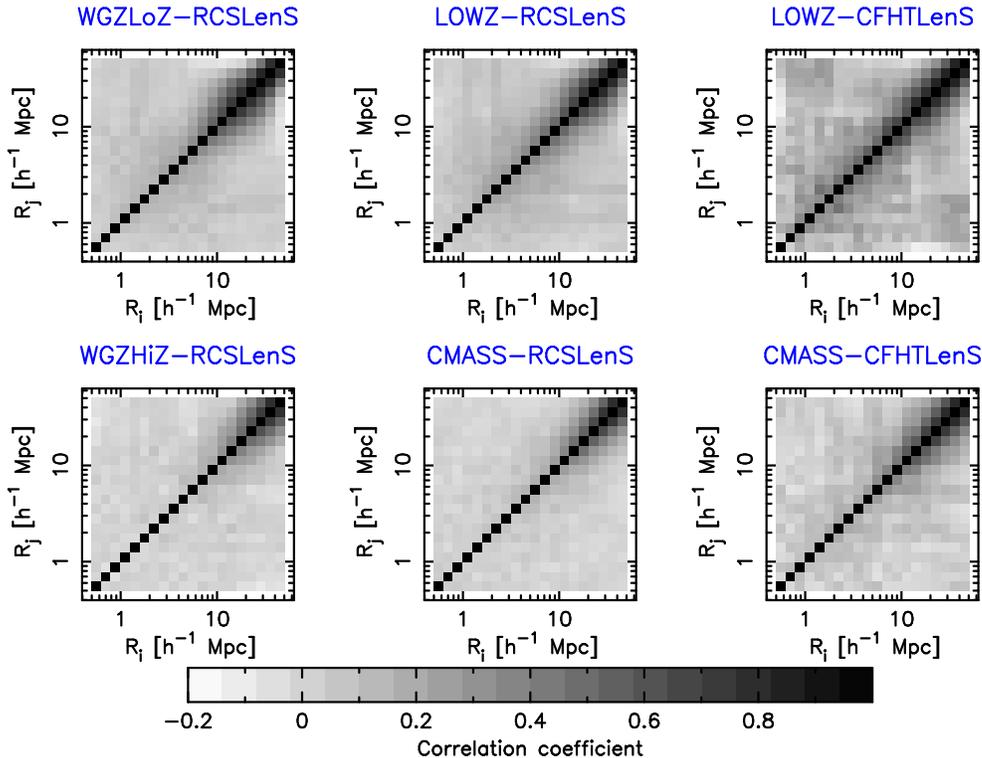}}}
\end{center}
\caption{Covariance matrices of the $\Delta \Sigma(R)$ measurements
  for different combinations of source-lens datasets, determined using
  the scatter across the 374 simulation lines-of-sight.  The
  covariance matrix $C_{ij}$ is displayed as a correlation matrix
  $C_{ij}/\sqrt{C_{ii} C_{jj}}$.}
\label{figdelsigcov}
\end{figure*}

We also constructed mock catalogues for each survey region including
the full survey mask of sources and lenses, implemented by stitching
together multiple simulation lines-of-sight and sub-sampling the
result to match the survey selection functions.  149 independent mock
catalogues for each survey region could be generated from the 374
simulations.  This masked simulation set allows the importance of the
survey selection function in the measurement error to be determined,
as described in the next sub-section.

\subsection{Determination of the covariance matrix}

We compared several techniques for obtaining errors in the measurement
of $\gamma_t(\theta)$ and $\Delta \Sigma(R)$:

\begin{itemize}

\item Data jack-knife errors, where 16 (4 $\times$ 4) jack-knife
  regions of typical dimension $\sim 2^\circ$ are used per lensing
  survey region, obtained by dividing the source distribution into
  sub-samples containing equal number of galaxies using constant
  R.A.\ and Dec.\ boundaries.

\item Simulated errors not including the survey mask.  We used the
  simulations discussed above, which comprise 374 lines-of-sight each
  covering 60 deg$^2$.  We measured the cross-correlations for each
  line-of-sight, and scaled the resulting scatter by $\sqrt{(60 \,
    {\rm deg}^2)/A_{\rm eff}}$ where $A_{\rm eff}$ is the effective
  (unmasked) area of each source region listed in Table \ref{tabdata}.

\item Simulated errors including the survey mask, implemented by
  stitching together multiple lines-of-sight for the simulations and
  sub-sampling the result to match the selection functions of the
  sources and the lenses, as discussed above.  We measured the
  cross-correlations for each of the resulting 149 mock catalogues per
  region; the standard deviation of these measurements constitutes our
  error estimate.

\item Jack-knife errors applied to both types of mock catalogues.  For
  the mocks without the survey mask, we scaled the resulting scatter
  by $\sqrt{(60 \, {\rm deg}^2)/A_{\rm eff}}$ for each region.

\end{itemize}

Figure \ref{figerrcomp} compares these error determinations for the
$\Delta \Sigma(R)$ statistic for the different combinations of
source-lens datasets, combining results in the different survey
regions.  The interpretation of these results is that on small scales,
the errors in the measurements are dominated by shape noise.  At
larger scales other effects become important, such as the same source
galaxies contributing to many stacks around different lenses, and the
``sample variance'' contributed by the particular realization of
large-scale structure within our lens samples.

All the different error estimates agree well for scales $R < 5 \,
h^{-1}$ Mpc for the low-redshift lenses $(0.15 < z < 0.43)$ and $R < 8
\, h^{-1}$ Mpc for the high-redshift lenses $(0.43 < z < 0.7)$,
following a scaling of the projected separation corresponding to a
given angular separation.  The jack-knife method applied to the
simulations produces an error estimate which matches on all scales
that obtained from the actual data.  The error determined from the
area-scaled unmasked simulations and fully-masked simulations matches
for all angles, suggesting that the survey mask is of secondary
importance for determining the measurement errors compared to the
total unmasked area.  However, these simulation errors are somewhat
lower than the jack-knife errors for larger separations.  We expect
the jack-knife errors to become unreliable for large scales as the
jack-knife regions become increasingly less independent.

As a result of these tests, we constructed our measurement errors
using the ensemble of $N_{\rm real} = 374$ mock catalogues without the
survey mask.  Figure \ref{figdelsigcov} illustrates the off-diagonal
elements of the resulting covariance matrix of the $\Delta\Sigma(R)$
statistic for the different combinations of source-lens datasets,
constructed as
\begin{eqnarray}
& & {\rm Cov}[\Delta\Sigma (R_i), \Delta\Sigma(R_j)] = \frac{1}{N_{\rm
      real}-1} \times \nonumber \\ & & \left[ \sum_{k=1}^{N_{\rm
        real}} \Delta\Sigma^k(R_i) \, \Delta\Sigma^k(R_j) -
    \overline{\Delta\Sigma(R_i)} \; \overline{\Delta\Sigma(R_j)}
    \right] ,
\end{eqnarray}
where $\Delta\Sigma^k$ is measured for the $k$th mock catalogue, and
$\overline{\Delta\Sigma(R_i)} \equiv \sum_{k=1}^{N_{\rm real}}
\Delta\Sigma^k(R_i) / N_{\rm real}$.  Significant correlations between
bins are evident for $R > 10 \, h^{-1}$ Mpc.

The inverse of these covariance matrices is used in the parameter fits
described below.  We correct the inverse covariance for the bias due
to its maximum-likelihood estimation (Hartlap et al.\ 2007) via
multiplication by the factor
\begin{equation}
\alpha = \frac{N_{\rm real} - N_{\rm bin} - 2}{N_{\rm real} - 1} ,
\end{equation}
where $N_{\rm bin}$ is the number of data bins used in the fit.  For
our analyses, $N_{\rm bin}/N_{\rm real} \approx 0.05$.

\section{Cosmological results}
\label{seccosmo}

\subsection{Projected galaxy auto-correlation function $w_p(R)$}
\label{secwpest}

\begin{figure}
\begin{center}
\resizebox{8cm}{!}{\rotatebox{270}{\includegraphics{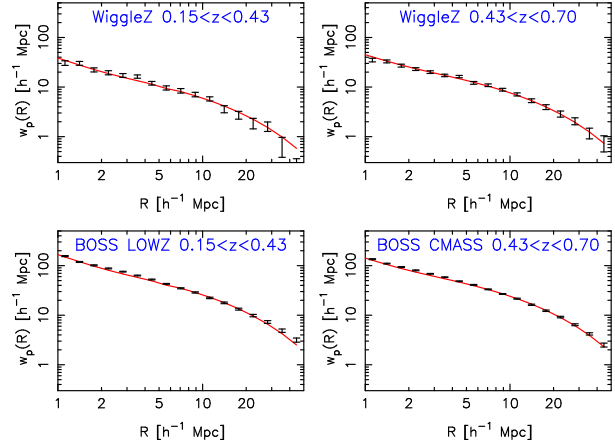}}}
\end{center}
\caption{The projected correlation function $w_p(R)$ for the different
  lens data samples used in our analysis.  Jack-knife errors are
  plotted, together with the best-fitting model using both the
  $w_p(R)$ and $\Delta \Sigma(R)$ measurements.}
\label{figwp}
\end{figure}

We measured the 2D galaxy correlation function $\xi_{gg}(R,\Pi)$ of
each lens sample, binning galaxy pairs by projected separation $R$ and
line-of-sight separation $\Pi$.  We hence determined the projected
correlation functions
\begin{equation}
w_p(R) = 2 \, \sum_{{\rm bins} \, i} \xi_{gg}(R,\Pi_i) \, \Delta \Pi_i ,
\end{equation}
where we summed over 10 logarithmically-spaced bins in $\Pi$ from
$0.1$ to $60 \, h^{-1}$ Mpc.  The measurements of $w_p(R)$ in 20
logarithmically-spaced bins in $R$ from $0.5$ to $50 \, h^{-1}$ Mpc
for the different lens samples are plotted in Figure \ref{figwp},
including jack-knife errors.  These projected correlation function
measurements are used for simple parameter fits in Section
\ref{secparfit}, and to determine the $E_G$ statistic in Section
\ref{seceg}.

\subsection{Measurements of $\sigma_8$}
\label{secparfit}

\begin{figure}
\begin{center}
\resizebox{8cm}{!}{\rotatebox{270}{\includegraphics{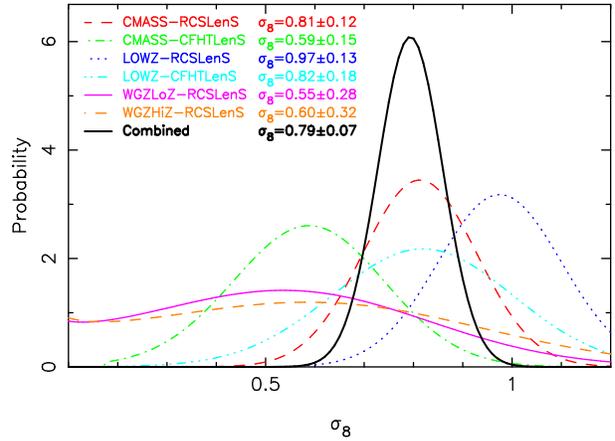}}}
\end{center}
\caption{The posterior probability distributions of the $\sigma_8$
  parameter, after marginalizing over the bias factors of the lens
  galaxies, for the different combinations of source-lens datasets.
  We also show the probability distribution of the fit to all
  datasets.}
\label{figprob}
\end{figure}

As an initial consistency test of the best-fitting cosmological
parameters in the $\Lambda$CDM model, we fitted the measurements of
$\Delta\Sigma(R)$ and $w_p(R)$ for each source-lens combination
varying just $\sigma_8$ and the galaxy bias of each lens sample,
fixing the other cosmological parameters at the values used to
construct the N-body simulations listed in Section \ref{secsim}.
Given that $\Delta\Sigma \propto b \, \sigma_8^2$ and $w_p \propto b^2
\, \sigma_8^2$, the degeneracy between these normalization parameters
is broken and they can be separately determined.  Fits were performed
using the full covariance matrix of $\Delta\Sigma(R)$ determined from
the mock catalogues, and a diagonal error matrix for $w_p(R)$ using
jack-knife errors.  The use of the latter is not significant: the
signal-to-noise of $w_p(R)$ is much higher than that of
$\Delta\Sigma(R)$, and the limiting factor for the final parameter
error is the noise in $\Delta\Sigma$, given the initial degeneracy
between $b$ and $\sigma_8$.

\begin{figure*}
\begin{center}
\resizebox{14cm}{!}{\rotatebox{270}{\includegraphics{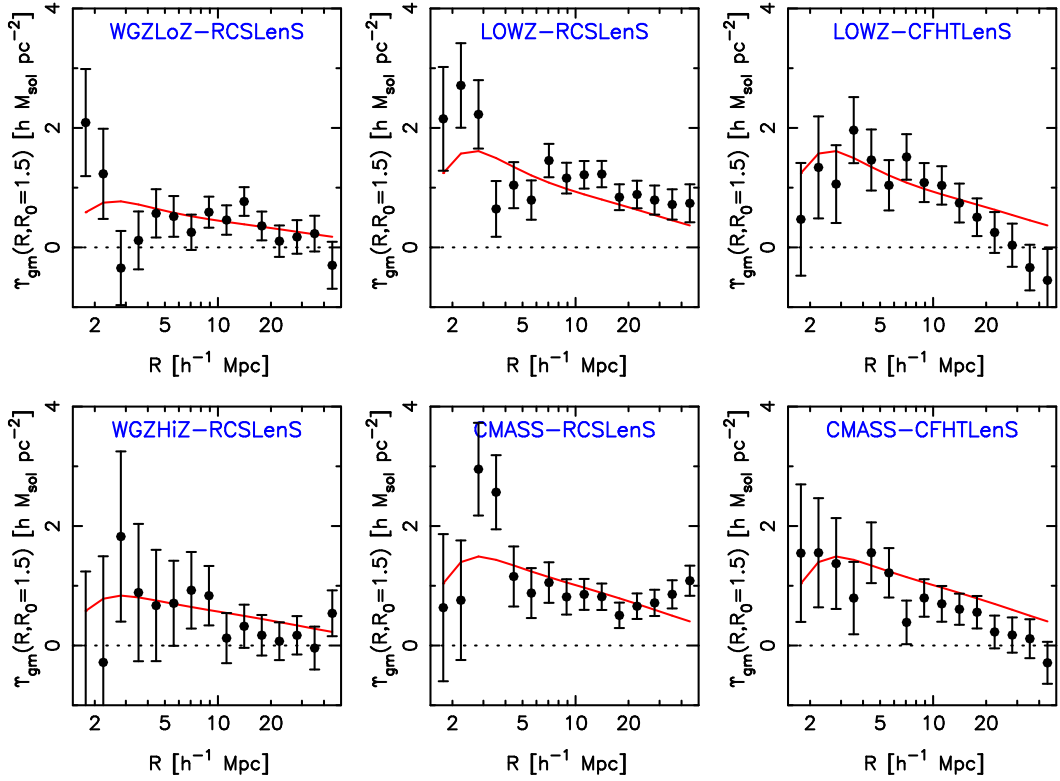}}}
\end{center}
\caption{The annular differential surface density statistic for the
  galaxy-mass cross-correlation, $\Upsilon_{gm}(R,R_0)$, measured for
  the different combinations of lens-source datasets assuming $R_0 =
  1.5 \, h^{-1}$ Mpc.  We also plot the best-fitting model for each
  cross-correlation using both the $w_p(R)$ and $\Delta \Sigma(R)$
  measurements.  The errors are based on measurements for a set of 374
  mock catalogues.  The horizontal dotted line marks $\Upsilon_{gm} =
  0$.}
\label{figups}
\end{figure*}

We calculated model predictions using Equation \ref{eqxigm} for
$\Delta\Sigma(R)$ and Equation \ref{eqwpmod} for $w_p(R)$ and fit to
the measurements over the range of scales $R > 5 \, h^{-1}$ Mpc
(noting that our results do not depend significantly on the choice of
minimum fitted scale).  The best-fitting models to the data are
overplotted in Figures \ref{figwp} and \ref{figups}.  The value of the
chi-squared statistic for the best-fitting model is $\chi^2 = 128.3$
for 113 degrees of freedom, indicating a good fit.

\begin{figure*}
\begin{center}
\resizebox{14cm}{!}{\rotatebox{270}{\includegraphics{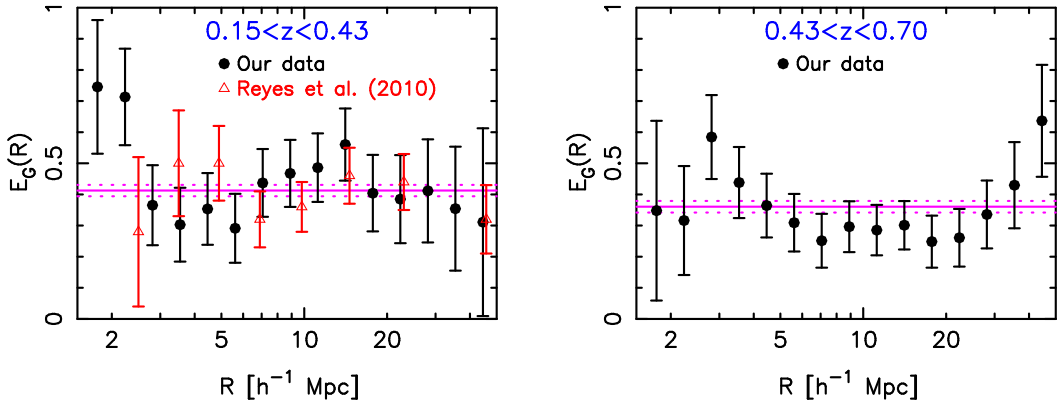}}}
\end{center}
\caption{$E_G(R)$ measurements in two independent redshift bins $0.15
  < z < 0.43$ and $0.43 < z < 0.7$, after combining the results from
  the different cross-correlations.  In the former case, the
  measurements of Reyes et al.\ (2010) are plotted as the open circles
  for comparison.  The horizontal solid lines are the prediction of
  standard gravity, $E_G = \Omega_m/f$, for our fiducial model
  $\Omega_m = 0.27$.  The horizontal dotted lines indicate the
  1-$\sigma$ variation that would result given $\Delta \Omega_m =
  0.02$, which is indicative of both the WMAP and {\it Planck} error
  in determining this parameter.  We note that the data points are
  correlated, with a covariance matrix displayed in Figure
  \ref{figegcov}.}
\label{figeg}
\end{figure*}

The combined measurement is $\sigma_8 = 0.79 \pm 0.07$ consistent with
the latest determinations from the {\it Planck} satellite ({\it
  Planck} collaboration 2015b).  Combining the separate fits to
determine the normalizations in two independent redshift bins, $0.15 <
z < 0.43$ (WGZLoZ, LOWZ) and $0.43 < z < 0.7$ (WGZHiZ, CMASS), we
determine $\sigma_8(z=0.32) = 0.75 \pm 0.08$ and $\sigma_8(z=0.57) =
0.54 \pm 0.07$.

In Figure \ref{figprob} we display the posterior probability
distributions of $\sigma_8$ for each source-lens dataset, marginalized
over the bias factors.  The datasets offer roughly comparable
constraining power, with CMASS-RCSLenS and WGZHiZ-RCSLenS producing
the most and least accurate determinations, respectively.

As a cross-check of the methodology we performed the same fits to the
$\Delta\Sigma(R)$ measurements from the mock catalogues for all the
combinations of source-lens datasets, using the full-survey
realizations including masks.  The average parameter measurement
across the realizations is $\sigma_8 = 0.80 \pm 0.03$ with average
value of $\chi^2/{\rm dof} = 50.5/47$, compared to the input parameter
value $\sigma_8 = 0.826$.  The slight offset of the fit to lower
values than the input is due to the artificial reduction in the
clustering amplitude of high-bias mocks constructed via Equation
\ref{eqbg}, as discussed in Section \ref{secsim}.  For $b=1$ mocks we
recover the input cosmology within the statistical error in the mean.

Future work will perform a full cosmological parameter analysis of
these lensing and clustering datasets, in combination with the CMB.

\subsection{Measurement of gravitational slip $E_G(R)$}
\label{seceg}

In this Section we use the measured galaxy-galaxy lensing
cross-correlations, in combination with the clustering strength and
redshift-space distortion properties of the lenses, to compute the
scale-dependent $E_G$ statistic defined in Equation \ref{eqeg} and
hence carry out a test of gravitational physics.  We calculated
$E_G(R)$ for each different combination of source-lens datasets, and
then combined the $E_G$ measurements in two independent redshift bins,
$0.15 < z < 0.43$ (WGZLoZ, LOWZ) and $0.43 < z < 0.7$ (WGZHiZ, CMASS).

Firstly we converted the measurements of $\Delta \Sigma(R)$ and
$w_p(R)$ for each source-lens combination into the annular
differential statistics $\Upsilon_{gm}(R)$ and $\Upsilon_{gg}(R)$
defined by Equations \ref{equpsgm} and \ref{equpsgg}, respectively.
We determined the values of $\Delta \Sigma(R_0)$ and $w_p(R_0)$ via a
power-law fit to the appropriate statistic, taking the fitting range
as $R_0/3 < R < 3R_0$.  Following Reyes et al.\ (2010) we assumed a
fiducial value $R_0 = 1.5 \, h^{-1}$ Mpc for our analysis, although we
also consider a range of other choices between $1$ and $3 \, h^{-1}$
Mpc.  We propagated the errors in $\Delta \Sigma(R_0)$ and $w_p(R_0)$
into the errors in $\Upsilon_{gm}(R)$ and $\Upsilon_{gg}(R)$, although
this source of error is negligible except when $R \approx R_0$.  We
determined the integral appearing in the first term of Equation
\ref{equpsgg} using a spline fit to the measured $w_p(R)$ (noting that
these details make a negligible contribution to error propagation).
Measurements of $\Upsilon_{gm}(R,R_0=1.5 \, h^{-1} {\rm Mpc})$ are
displayed in Figure \ref{figups} for the different source and lens
combinations.

We cross-checked the errors in the determination of
$\Upsilon_{gm}(R)$ by repeating the procedure for the ensemble of mock
catalogues.  The scatter in the measurements across the mocks agreed
with the propagated value of the error within a few per cent.

We then determined $E_G(R)$ for the different samples, using the
values of $\beta$ quoted in the respective redshift-space distortion
analyses of the lenses, as reproduced in Table \ref{tabeg}.  For the
BOSS galaxies we used $\beta_{\rm LOWZ} = 0.38 \pm 0.11$ and
$\beta_{\rm CMASS} = 0.36 \pm 0.06$ (Sanchez et al.\ 2014).  We
performed new RSD fits to the WiggleZ data for redshift ranges $0.15 <
z < 0.43$ and $0.43 < z < 0.7$ following the procedures of Blake et
al.\ (2011), obtaining $\beta_{\rm WGZLoZ} = 0.66 \pm 0.10$ and
$\beta_{\rm WGZHiZ} = 0.63 \pm 0.08$.  We propagated the errors in
$\Upsilon_{gm}$, $\Upsilon_{gg}$ and $\beta$ into the error in $E_G$.
We then combined the different measurements of $E_G(R)$ in the two
redshift bins $0.15 < z < 0.43$ and $0.43 < z < 0.7$, using
inverse-variance weighting.

\begin{figure}
\begin{center}
\resizebox{8cm}{!}{\rotatebox{270}{\includegraphics{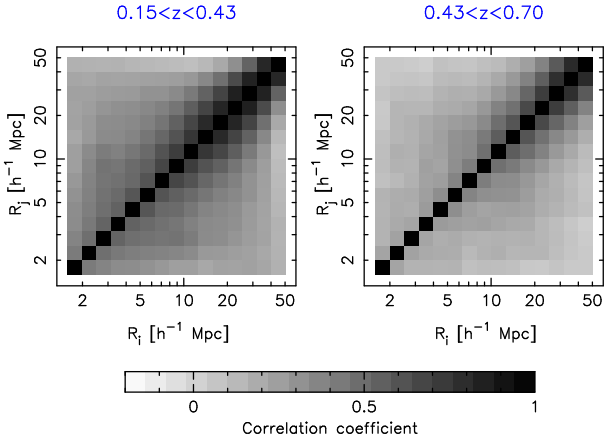}}}
\end{center}
\caption{Covariance matrix of the $E_G(R)$ measurements in the two
  redshift bins, displayed as a correlation matrix
  $C_{ij}/\sqrt{C_{ii} C_{jj}}$.}
\label{figegcov}
\end{figure}

The final results are displayed in Figure \ref{figeg}, corresponding
to a $\approx 20\%$ determination of $E_G$ at each scale and redshift.
We overplot the GR model prediction at each redshift, $E_G =
\Omega_m/f(z)$ for our fiducial model $\Omega_m = 0.27$ (together with
its 1-$\sigma$ variation give the typical error in determination of
this parameter from the CMB, $\Delta \Omega_m = 0.02$), and the
measurements of Reyes et al.\ (2010), which correspond to the
lower-redshift bin.  The covariance matrix of these $E_G$ measurements
is shown in Figure \ref{figegcov}.

Using the covariance matrix, we fit a model of constant $E_G(R)$ to
the data, comparing fits over the ranges $R > 2 R_0 = 3 \, h^{-1}$ Mpc
and $R > 10 \, h^{-1}$ Mpc.  The resulting values of $\overline{E_G}$
for each individual source-lens dataset are listed in Table
\ref{tabeg}.  The combined measurements in the two unique redshift
bins $z = (0.32, 0.57)$ are $(0.40 \pm 0.09, 0.31 \pm 0.06)$ for $R >
3 \, h^{-1}$ Mpc and $(0.48 \pm 0.10, 0.30 \pm 0.07)$ for $R > 10 \,
h^{-1}$ Mpc, and hence are not significantly sensitive to the minimum
scale used.  The $\chi^2$ of the best-fitting constant $E_G$ model to
$R > 10 \, h^{-1}$ Mpc is $(5.5, 5.6)$ in the two redshift bins, for 6
degrees of freedom.  The consistency of these measurements with the
predictions of the perturbed GR model demonstrates that this model has
successfully predicted both the scale-independence of the signal, and
the overall amplitude.

Our measurement agrees with the fit obtained by Reyes et al.\ for
$z=0.32$, $\overline{E_G} = 0.39 \pm 0.07$.  The errors obtained in
the value of $\overline{E_G}$ are also similar; the Reyes et
al.\ measurements are based on SDSS imaging data which covers much
wider area, but provides a significantly lower density of measured
shapes.

\begin{figure}
\begin{center}
\resizebox{8cm}{!}{\rotatebox{270}{\includegraphics{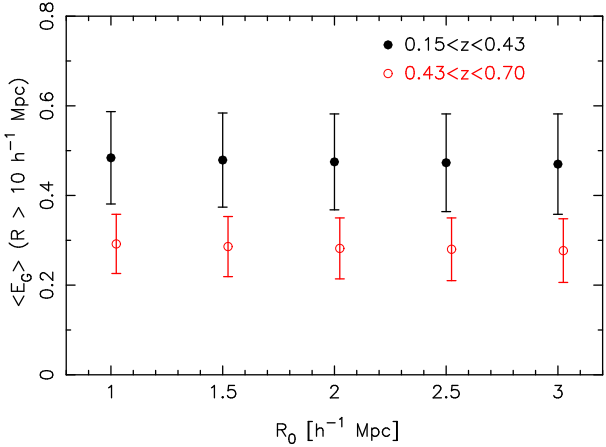}}}
\end{center}
\caption{The dependence of the fit of constant $\overline{E_G}$ to the
  range of scales $R > 10 \, h^{-1}$ Mpc on the value of $R_0$ chosen
  for the small-scale suppression in the annular differential surface
  density statistics.}
\label{figegr0}
\end{figure}

\begin{figure}
\begin{center}
\resizebox{8cm}{!}{\rotatebox{270}{\includegraphics{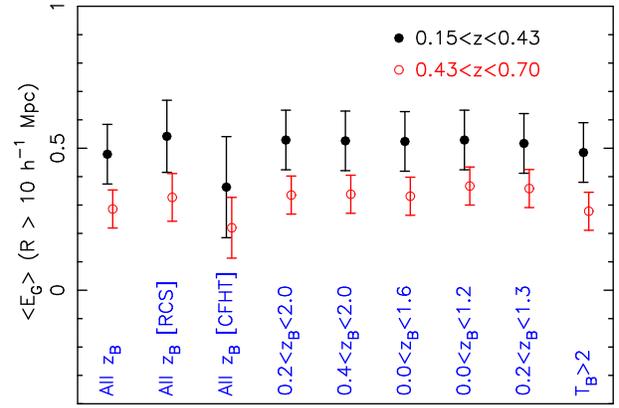}}}
\end{center}
\caption{The dependence of the fit of constant $\overline{E_G}$ to the
  range of scales $R > 10 \, h^{-1}$ Mpc on the range of photometric
  redshifts allowed in the analysis.}
\label{figegzphot}
\end{figure}

We now test the sensitivity of our measurements to two important
analysis choices.  First, in Figure \ref{figegr0} we show the results
of repeating these fits of $\overline{E_G}$ to the range $R > 10 \,
h^{-1}$ Mpc, in the two redshift bins, for different choices of the
parameter $R_0$ used for small-scale suppression when measuring
$\Upsilon_{gm}$ and $\Upsilon_{gg}$, in the range $1 < R_0 < 3 \,
h^{-1}$ Mpc.  These results show that our measurement of
$\overline{E_G}$ is insensitive to the choice of $R_0$.  Secondly, in
Figure \ref{figegzphot} we compare the fiducial measurements to
repeated fits applying different cuts in the range of photometric
redshifts included in the analysis: $0.2 < z_B < 2.0$, $0.4 < z_B <
2.0$, $0 < z_B < 1.6$, $0 < z_B < 1.2$ and $0.2 < z_B < 1.3$.  We also
show results just including RCSLenS or CFHTLenS regions, and for a
``blue'' sub-sample selected by a cut $T_B > 2$ (which will be subject
to a lower level of intrinsic alignments, as discussed in Section
\ref{secsysalign}).  The measured values of $\overline{E_G}$ never
differ by a significant amount from the fiducial case.

Finally, we assess the significance of any bias in the mean or
standard deviation of $E_G$ induced by non-Gaussianity in its
posterior probability distribution.  Since it is a ratio of noisy
quantities, the error distribution of $E_G$ is not necessarily
Gaussian, even if the numerator and denominator are
Gaussian-distributed.  We compared the mean and standard deviation
obtained from direct error propagation with the distributions arising
from Monte-Carlo sampling the relevant quantities in the numerator and
denominator.  We found that, averaging over bins with $R > 10 \,
h^{-1}$ Mpc, the mean of $E_G$ was biased low by $\Delta E_G = (0.03,
0.01)$ in the two redshift ranges $z=(0.32, 0.57)$, which is just
under half the statistical error for the lower-redshift bin (the
greater impact in the lower-redshift bin is driven by the greater
fractional error in $\beta$).  The width of the 1-$\sigma$ and
2-$\sigma$ confidence regions agreed within $5\%$ for the two cases.
We conclude that systematic errors due to skewness in the $E_G$
distribution are sub-dominant to statistical errors.

An additional issue would arise if the numerator and denominator of
the $E_G$ relation are additionally correlated, since then the mean
(or median or mode) of the ratio would not be an unbiased estimator of
the true value (e.g.\ Viola et al. 2014, Marian et al.\ 2015).  We
have estimated the correlation strength between the galaxy clustering
and the galaxy-galaxy lensing correlation functions using Gaussian
analytic covariances and found the correlation to be negligible. This
is driven by two effects: (a) the samples studied in this work are
dominated by shot noise rather than cosmic variance, which tends to
de-correlate two-point statistics, and (b) $w_p$ and $\Delta \Sigma$
are related to the underlying matter power spectrum via different
weighting functions, such that the same separations in $w_p$ and
$\Delta \Sigma$ are sensitive to different ranges of $P_{mm}(k)$.

\begin{table*}
\caption{Values of constant $E_G$ fit to the measurements for each
  combination of source-lens datasets, using the range of scales $R >
  3 \, h^{-1}$ Mpc and $R > 10 \, h^{-1}$ Mpc.  Best-fitting values of
  $\chi^2$ are shown for each case, and the measurements of the input
  RSD parameter $\beta$.}
\begin{center}
\begin{tabular}{cccccc}
\hline
Survey & $\beta$ & $\overline{E_G}(R > 3 \, h^{-1}{\rm Mpc})$ &
$\chi^2$/dof & $\overline{E_G}(R > 10 \, h^{-1}{\rm Mpc})$ &
$\chi^2$/dof \\
\hline
WGZLoZ-RCSLenS & $0.65 \pm 0.10$ & $0.36 \pm 0.15$ & $16.5/11$ & $0.52 \pm 0.21$ & $13.3/6$ \\
LOWZ-RCSLenS   & $0.38 \pm 0.11$ & $0.40 \pm 0.14$ & $13.6/11$ & $0.54 \pm 0.15$ & $ 4.1/6$ \\
LOWZ-CFHTLenS  & $0.38 \pm 0.11$ & $0.41 \pm 0.16$ & $ 8.4/11$ & $0.36 \pm 0.18$ & $ 4.9/6$ \\
WGZHiZ-RCSLenS & $0.63 \pm 0.08$ & $0.38 \pm 0.15$ & $10.0/11$ & $0.27 \pm 0.17$ & $ 8.1/6$ \\
CMASS-RCSLenS  & $0.36 \pm 0.06$ & $0.36 \pm 0.08$ & $18.9/11$ & $0.36 \pm 0.09$ & $13.3/6$ \\
CMASS-CFHTLenS & $0.36 \pm 0.06$ & $0.24 \pm 0.09$ & $10.8/11$ & $0.23 \pm 0.11$ & $ 5.9/6$ \\
\hline
Combined $0.15<z<0.43$ & - & $0.40 \pm 0.09$ & $10.3/11$ & $0.48 \pm 0.10$ & $ 5.5/6$ \\
Combined $0.43<z<0.70$ & - & $0.31 \pm 0.06$ & $ 8.1/11$ & $0.30 \pm 0.07$ & $ 5.6/6$ \\
\hline
\end{tabular}
\end{center}
\label{tabeg}
\end{table*}

\section{Conclusions}
\label{secconc}

In this study we have performed a new test of gravitational physics
based on a consistency check between the amplitude of peculiar
velocities, measured via redshift-space distortion in galaxy samples
from the WiggleZ and BOSS surveys, and the galaxy-galaxy lensing
signal imprinted in the shapes of background galaxies in the CFHTLenS
and RCSLenS datasets by density fluctuations traced by these lenses.
Our results agree with the predictions of GR for a perturbed FRW
metric, in a flat $\Lambda$CDM Universe with matter density consistent
with observations of the CMB.

In particular we produce new measurements of the ``gravitational
slip'' statistic $E_G = \Upsilon_{gm}/(\beta \, \Upsilon_{gg})$,
estimating values $0.48 \pm 0.10$ at $z=0.32$ and $0.30 \pm 0.07$ at
$z=0.57$ when averaging over scales $10 < R < 50 \, h^{-1}$ Mpc,
compared to model predictions of $E_G = 0.41$ and $0.36$,
respectively.  The results are consistent with those obtained when
averaging over $3 < R < 50 \, h^{-1}$ Mpc.  When carrying out this
measurement we suppressed small-scale information from $R < R_0 = 1.5
\, h^{-1}$ Mpc using the annular differential surface density
statistic, although we find that our results are in fact insensitive
to the choice of $R_0$.

Assuming a $\Lambda$CDM cosmological model and fixing all the
parameters apart from $\sigma_8$, we determine $\sigma_8(z=0.32) =
0.75 \pm 0.08$ and $\sigma_8(z=0.57) = 0.54 \pm 0.07$ by fitting to
the differential surface density $\Delta\Sigma$ and projected lens
correlation function $w_p$, after marginalizing over separate linear
bias factors of each lens sample.  These results are also consistent
with the expected growth of structure in a perturbed $\Lambda$CDM
Universe.

In terms of methodology, we particularly note the systematic bias that
can occur in the measurement of $\Delta \Sigma(R)$ when using
photo-$z$ information to select source-lens pairs, and how this can be
reduced by using the full redshift probability of each source in the
estimator.  We find that our results are not sensitive to the range of
photometric redshifts included in the analysis.

We note that a measurement of $E_G$ using lensing of the Cosmic
Microwave Background by BOSS-CMASS galaxies was recently presented by
Pullen et al.\ (2015), determining $E_G(z=0.57) = 0.243 \pm 0.060 {\rm
  (stat)} \pm 0.013 {\rm (sys)}$.  This measurement is consistent with
ours, with both results lying below the standard model prediction.

Combinations of gravitational lensing and redshift-space distortion
data offer a powerful opportunity to test large-scale gravitational
physics in the search for an understanding of the physical nature of
dark energy.  Datasets such as CFHTLenS, RCSLenS, the Kilo-Degree
Survey (KiDS), Dark Energy Survey (DES) and the Hyper-Suprime Cam
(HSC) lensing survey, promise that the constraining power of such
tests will continue to improve rapidly and provide increasingly
stringent tests of fundamental cosmology.

\section*{Acknowledgments}

We thank the anonymous referee for providing useful comments on the
paper.

We are grateful to the RCS2 team for planning the survey, applying for
observing time, and conducting the observations. We acknowledge use of
the Canadian Astronomy Data Centre operated by the Dominion
Astrophysical Observatory for the National Research Council of
Canada’s Herzberg Institute of Astrophysics.  We would like to thank
Matthias Bartelmann for being our external blinder, revealing which of
the four catalogues analysed was the true unblinded catalogue at the
end of this study.

CB acknowledges the support of the Australian Research Council through
the award of a Future Fellowship, and thanks the Department of Physics
and Astronomy at the University of Canterbury, Christchurch, New
Zealand for their kind hospitality during the development of this
paper.  We also acknowledge the Aspen Center for Physics (NSF grant
1066293) where some of this work took place.

Part of this work was performed using the SwinSTAR supercomputer at
Swinburne University of Technology, and CB is grateful to Jarrod
Hurley and the HPC support team for invaluable technical help during
this period.

CH and AC acknowledge support from the European Research Council under
the EC FP7 grant number 240185.  TE is supported by the Deutsche
Forschungsgemeinschaft in the framework of the TR33 `The Dark
Universe'.  JHD is supported by the NSERC of Canada.  BJ acknowledges
support by an STFC Ernest Rutherford Fellowship, grant reference
ST/J004421/1.  HH is supported by an Emmy Noether grant (No. Hi
1495/2-1) of the Deutsche Forschungsgemeinschaft.  MV acknowledges
support from the European Research Council under FP7 grant number
279396 and the Netherlands Organisation for Scientific Research (NWO)
through grants 614.001.103.

Computations for the $N$-body simulations were performed on the GPC
supercomputer at the SciNet HPC Consortium. SciNet is funded by: the
Canada Foundation for Innovation under the auspices of Compute Canada;
the Government of Ontario; Ontario Research Fund - Research
Excellence; and the University of Toronto.

The WiggleZ Dark Energy Survey received financial support from the
Australian Research Council Discovery Project program.  We acknowledge
the dedicated work of the staff of the Australian Astronomical
Observatory in the development and support of the AAOmega
spectrograph, and the running of the AAT.

Funding for SDSS-III has been provided by the Alfred P.\ Sloan
Foundation, the Participating Institutions, the National Science
Foundation, and the U.S.\ Department of Energy Office of Science. The
SDSS-III web site is {\tt http://www.sdss3.org/}.

SDSS-III is managed by the Astrophysical Research Consortium for the
Participating Institutions of the SDSS-III Collaboration including the
University of Arizona, the Brazilian Participation Group, Brookhaven
National Laboratory, Carnegie Mellon University, University of
Florida, the French Participation Group, the German Participation
Group, Harvard University, the Instituto de Astrofisica de Canarias,
the Michigan State/Notre Dame/JINA Participation Group, Johns Hopkins
University, Lawrence Berkeley National Laboratory, Max Planck
Institute for Astrophysics, Max Planck Institute for Extraterrestrial
Physics, New Mexico State University, New York University, Ohio State
University, Pennsylvania State University, University of Portsmouth,
Princeton University, the Spanish Participation Group, University of
Tokyo, University of Utah, Vanderbilt University, University of
Virginia, University of Washington, and Yale University.

{\it Author contributions:} All authors contributed to the development
and writing of this paper.  The authorship list reflects the lead
authors of this paper (CB, SJ and CH) followed by an alphabetical
group.  This group includes key contributors to the science analysis
and interpretation in this paper, the founding core team of RCSLenS,
and those whose long-term significant effort produced the RCSLenS data
product.  HH led the RCSLenS collaboration.

\appendix

\section{Calibration corrections and systematics tests}

\subsection{Systematics tests manipulating source shears}
\label{secsysshear}

\begin{figure*}
\begin{center}
\resizebox{14cm}{!}{\rotatebox{0}{\includegraphics{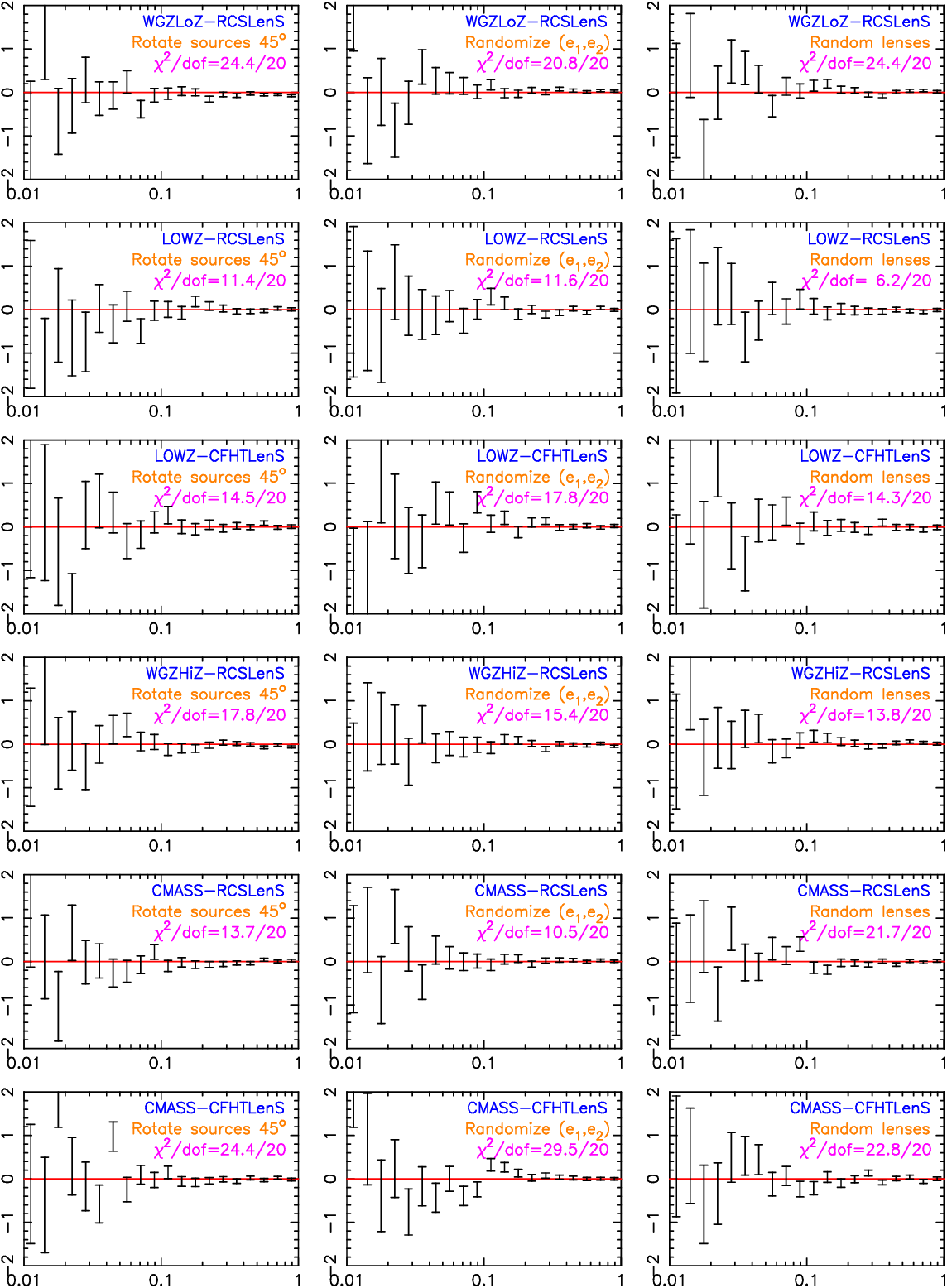}}}
\end{center}
\caption{Systematics tests of the $\gamma_t(\theta)$ measurement
  applied to the galaxy-galaxy lensing cross-correlations for the
  different combinations of source-lens datasets.  The $x$-axis is the
  angular separation $\theta$ in degrees, the $y$-axis plots $10^5
  \gamma_t$.  The left, middle, and right columns respectively show
  measurements following rotation of sources by $45^\circ$,
  randomization of shear values, and replacement of the lens catalogue
  by a random catalogue.  $\chi^2$ statistics are quoted for the
  measurements with respect to a model of zero, indicated by the
  horizontal line.}
\label{figsysshear}
\end{figure*}

We performed a series of systematics tests manipulating the source and
lens catalogues and re-measuring the average tangential shear
$\gamma_t(\theta)$:
\begin{itemize}
\item We rotated the sources by $45^\circ$ (i.e., $e_{1,{\rm new}} =
  e_{2,{\rm old}}$ and $e_{2,{\rm new}} = - e_{1,{\rm old}}$).
\item We randomized the shears (i.e., randomly shifted each pair of
  values $(e_1,e_2)$ to a different source galaxy).
\item We replaced the lens catalogue by a random catalogue.
\end{itemize}
The results of these tests are shown in Figure \ref{figsysshear}; in
all cases we should find $\gamma_t(\theta) = 0$ in the absence of
systematic errors.  The covariance matrices for each measurement are
obtained by applying the same systematics-test operation to each mock
catalogue; this covariance allows us to evaluate the $\chi^2$
statistic of the $\gamma_t = 0$ model in each case.  These values are
printed in each panel along with the number of degrees of freedom,
${\rm dof} = 20$, and indicate consistency with the model.

\subsection{Shear bias calibration corrections}
\label{secsyscal}

\begin{figure}
\begin{center}
\resizebox{8cm}{!}{\rotatebox{0}{\includegraphics{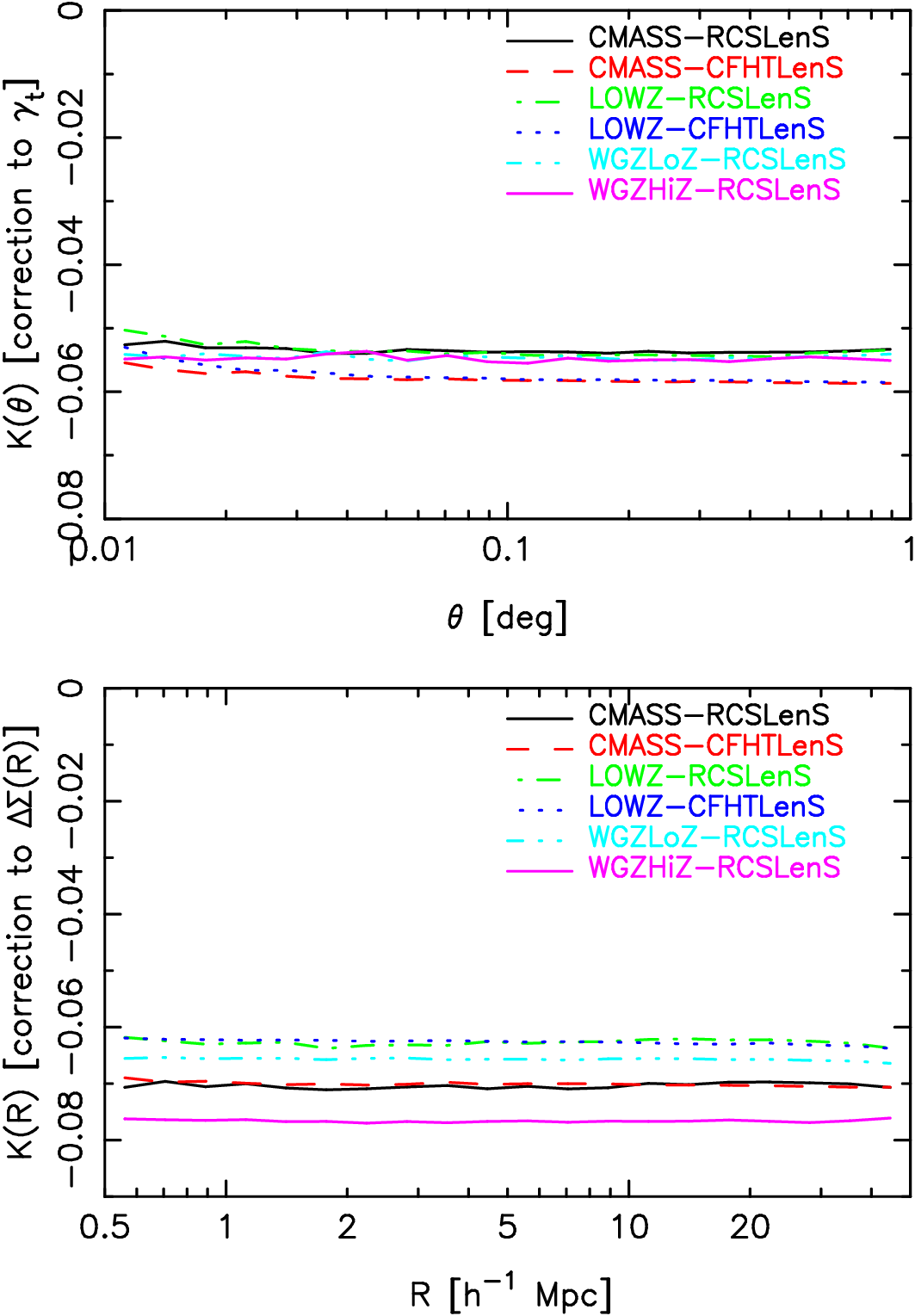}}}
\end{center}
\caption{Multiplicative shear bias corrections to be applied to
  $\gamma_t(\theta)$ (upper panel) and $\Delta \Sigma(R)$ (lower
  panel), for the different combinations of source-lens datasets.}
\label{figcorrmult}
\end{figure}

The effect of the additive shear calibration bias is negligible for
galaxy-galaxy lensing.  Figure \ref{figcorrmult} displays the
multiplicative shear bias corrections $K(\theta)$ and $K(R)$, defined
by Equations \ref{eqktheta} and \ref{eqkr}, which are applied to the
galaxy-galaxy lensing measurements by multiplying the estimated values
by $(1+K)^{-1}$.  Corrections are shown for each combination of
source-lens datasets, combining the different survey fields.  These
corrections are approximately independent of scale and have values $K
\approx -0.06$.  We note that the determinations of $K(R)$ depend on
the redshift distribution of the source and lens samples, through the
weighting factor $\Sigma_c^{-1}$, which imprints an extra variation
between datasets compared to $K(\theta)$.

\subsection{Photo-$z$ systematic test: SIS fits in $z_B$ slices}
\label{secsysphotz}

\begin{figure}
\begin{center}
\resizebox{8cm}{!}{\rotatebox{270}{\includegraphics{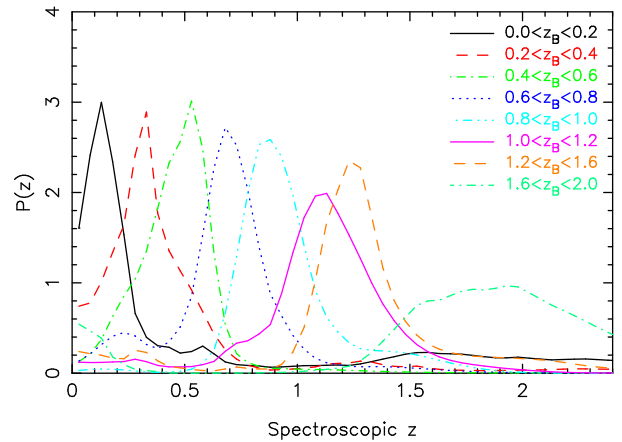}}}
\end{center}
\caption{Stacked RCSLenS {\small BPZ} redshift probability
  distributions, weighted by the {\it lens}fit weights, in a series of
  $z_B$ slices.}
\label{fignzphotz}
\end{figure}

Systematic errors in the {\small BPZ} photometric redshift probability
distributions would imprint errors in the determination of $\Delta
\Sigma(R)$, which relies on the computation of
$\overline{\Sigma_c^{-1}}$ for each source-lens pair from an integral
over redshift weighted by $p_{\rm BPZ}(z)$, and in the cosmological
modelling, which uses the source redshift probability distribution
$p_s(z)$ derived from stacking the individual $p_{\rm BPZ}(z)$
functions.  We tested for such systematics by computing the amplitude
of the tangential shear around a fixed set of foreground lenses in a
series of eight photometric redshift bins defined using $z_B$ --
$(0-0.2, 0.2-0.4, 0.4-0.6, 0.6-0.8, 0.8-1.0, 1.0-1.2, 1.2-1.6)$ -- in
order to determine whether this amplitude scaled with redshift in the
expected manner.  A convenient method for quantifying the results is
to assume a singular isothermal sphere (SIS) model for the lenses,
characterized by a velocity dispersion $\sigma_v$, and to verify that
the values of $\sigma_v$ derived from the shear of each photo-$z$
source slice are consistent.  The shear profile for SIS lenses is
given in terms of the Einstein radius $\theta_E$ by $\gamma_t(\theta)
= \theta_E/2\theta$, where
\begin{equation}
\theta_E = 4 \pi \left( \frac{\sigma_v}{c} \right)^2 \left<
\frac{D_{ls}}{D_s} \right> ,
\end{equation}
and the geometrical factor is given by
\begin{equation}
\left< \frac{D_{ls}}{D_s} \right> = \int_0^\infty dz \, p_l(z)
\int_z^\infty dz' \, p_s(z') \left[ \frac{\chi(z') -
    \chi(z)}{\chi(z')} \right] .
\end{equation}
Figure \ref{fignzphotz} displays the stacked RCSLenS {\small BPZ}
redshift probability distributions $p_s(z)$, weighted by the {\it
  lens}fit weights, in each of the photo-$z$ slices.  A signal is
produced even for low values of $z_B$, because sources are scattered
to apparent low redshifts from true higher redshifts.

\begin{figure}
\begin{center}
\resizebox{6cm}{!}{\rotatebox{0}{\includegraphics{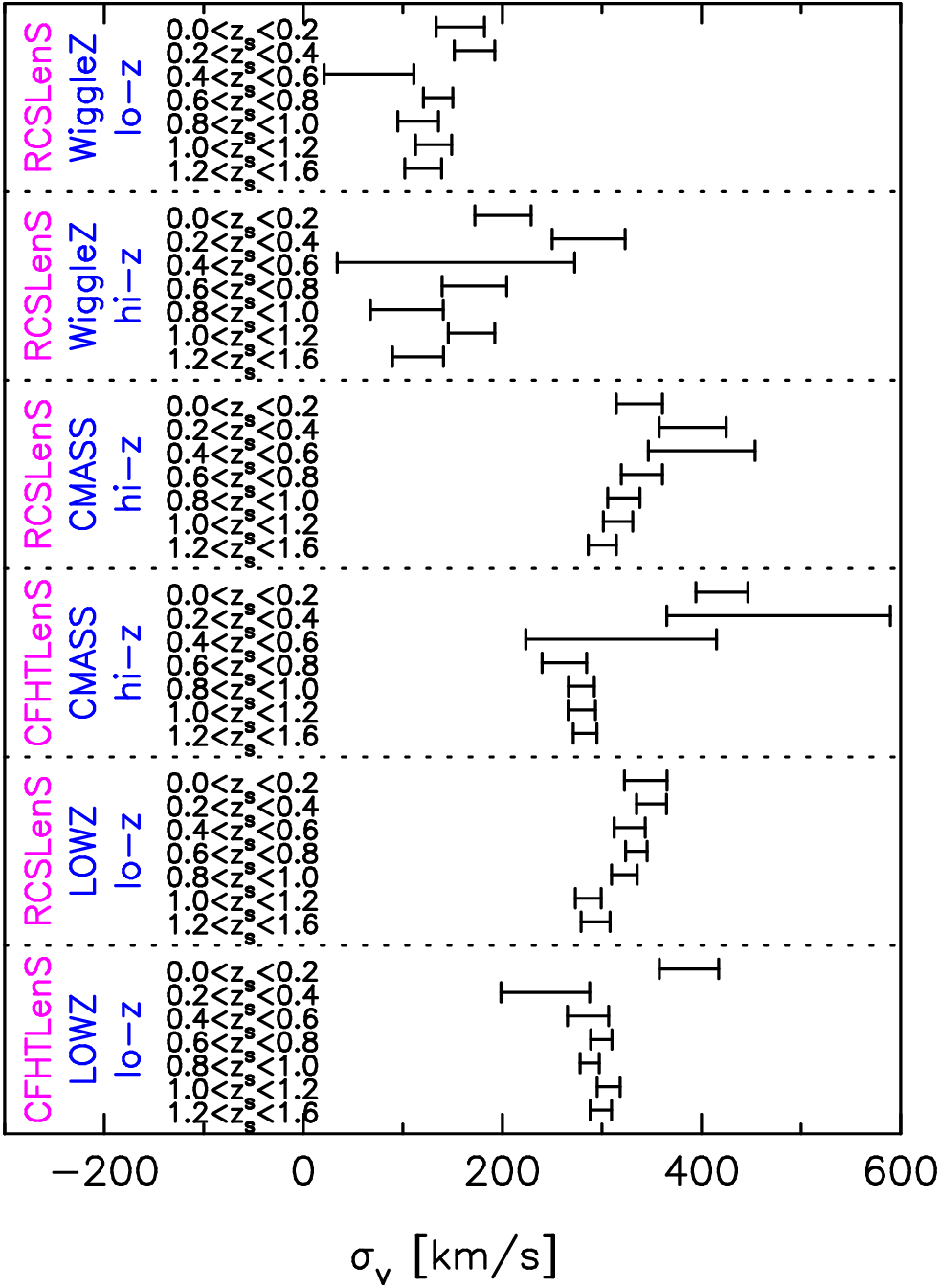}}}
\end{center}
\caption{The SIS velocity dispersion fit to the average tangential
  shear around a series of foreground lens samples, for different sets
  of sources in photometric redshift slices split by $z_B$.  The
  source redshift distributions for each slice, needed to model the
  resulting signal, are obtained from the stacked {\small BPZ}
  redshift probability distributions obtained for each source.}
\label{figsisphotz}
\end{figure}

Figure \ref{figsisphotz} displays the fits for the $\sigma_v$
parameter of the four different lens samples -- WGZLoZ, WGZHiZ, LOWZ
and CMASS -- in each photo-$z$ bin, using these $p(z_s)$ functions
determined from the outputs of {\small BPZ}.  The fits are performed
to $\gamma_t(\theta)$ measurements in the range $0.01^\circ < \theta <
1^\circ$.  The inferred values of $\sigma_v$ are generally consistent
as the source photo-$z$ slice changes, with the possible exception of
the lowest redshift slice $0 < z_B < 0.2$ for CFHTLenS, which contains
a negligible number of galaxies.

\subsection{Intrinsic alignments systematic test: SIS fits to $T_B$ samples}
\label{secsysalign}

\begin{figure}
\begin{center}
\resizebox{6cm}{!}{\rotatebox{0}{\includegraphics{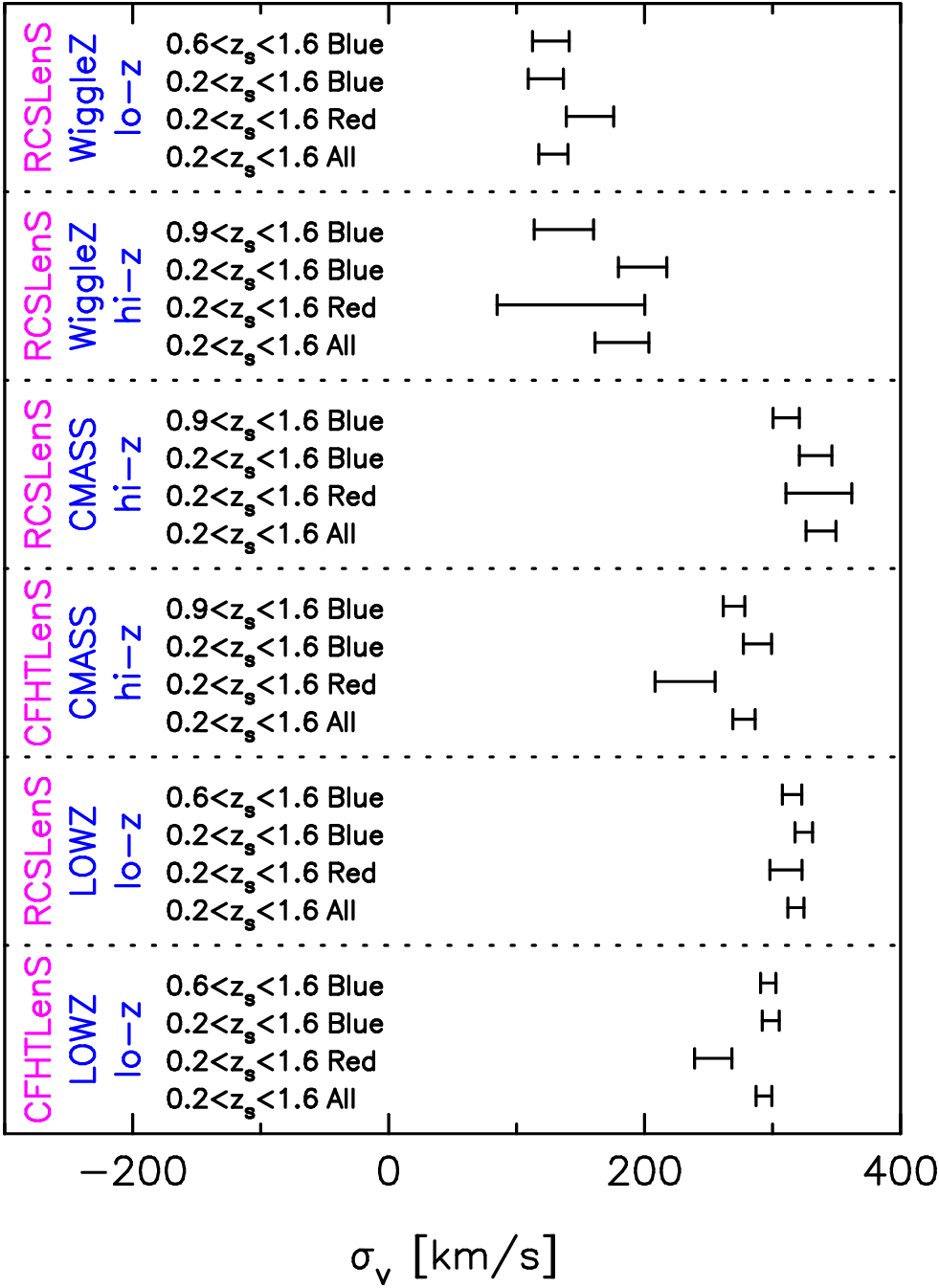}}}
\end{center}
\caption{The SIS velocity dispersion fit to the average tangential
  shear around a series of foreground lens samples, for different sets
  of sources split by galaxy type according to $T_B$.  The source
  redshift distributions for each slice, needed to model the resulting
  signal, are obtained from the stacked {\small BPZ} redshift
  probability distributions obtained for each source.}
\label{figsistb}
\end{figure}

Another potential systematic effect that could impact our measurements
is the intrinsic alignment of the background source sample with the
foreground lens distribution.  This effect is expected to
preferentially diminish the average tangential shear of red source
galaxies compared to blue galaxies.  The {\small BPZ} pipeline returns
a source galaxy type in the range $0 \le T_B \le 6$, and we used this
parameter to divide the sources into sub-samples of red ($T_B < 2$)
and blue ($T_B > 2$) galaxies.  Following Section \ref{secsysphotz},
we then determined the velocity dispersion of foreground lenses in the
SIS model for the blue, red and combined source samples in the range
$0.2 < z_B < 1.6$.  We also performed measurements for a blue sample
over a photo-$z$ range significantly higher than the maximum redshift
of the spectroscopic lenses ($0.9 < z_B < 1.6$ for WGZHiZ and CMASS,
and $0.6 < z_B < 1.6$ for WGZLoZ and LOWZ.  The source distributions
$p_s(z)$ for each case were determined by stacking the appropriate
subset of {\small BPZ} redshift distributions.  Results are displayed
in Figure \ref{figsistb}.  The inferred values of $\sigma_v$ are again
generally consistent amongst the samples, with the possible exception
of the cross-correlations between CFHTLenS and BOSS red galaxies.
However, as shown in Figure \ref{figegzphot}, the fitted value of
$E_G$ is not significantly affected by cutting the sources to a blue
sample.

\subsection{Source-lens association correction $B(R)$}
\label{secboost}

Some sources may be clustered or associated with the lenses, but
scattered to higher redshifts by photo-$z$ errors.  These sources will
not be lensed, diluting the cross-correlation signal.  The strength of
this effect may be determined by measuring the excess in the number
counts of source galaxies in the vicinity of lens galaxies, compared
to a random distribution of lenses.  The resulting multiplicative bias
in the measurement of $\Delta \Sigma(R)$ may be corrected by boosting
the signal by
\begin{equation}
B(R) = \frac{\sum_{{\rm sources} \, i} \sum_{{\rm data \, lenses} \,
    j} w^s_i \, w^l_j \, \Sigma_{c,ij}^{-1} \, \Theta(i,j)}{\sum_{{\rm
      sources} \, i} \sum_ {{\rm random \, lenses} \, j} w^s_i \,
  w^r_j \, \Sigma_{c,ij}^{-1} \, \Theta(i,j)} ,
\end{equation}
where $w^r$ denotes the weights of the random lenses (normalized such
that $\sum_j w^r_j = \sum_j w^l_j$), and where $\Sigma_{c,ij}^{-1}$
should be replaced by $\overline{\Sigma_c^{-1}}_{ij}$ when moving from
the estimator of Equation \ref{eqdelsigest2} to \ref{eqdelsigest1}.

\begin{figure*}
\begin{center}
\resizebox{14cm}{!}{\rotatebox{270}{\includegraphics{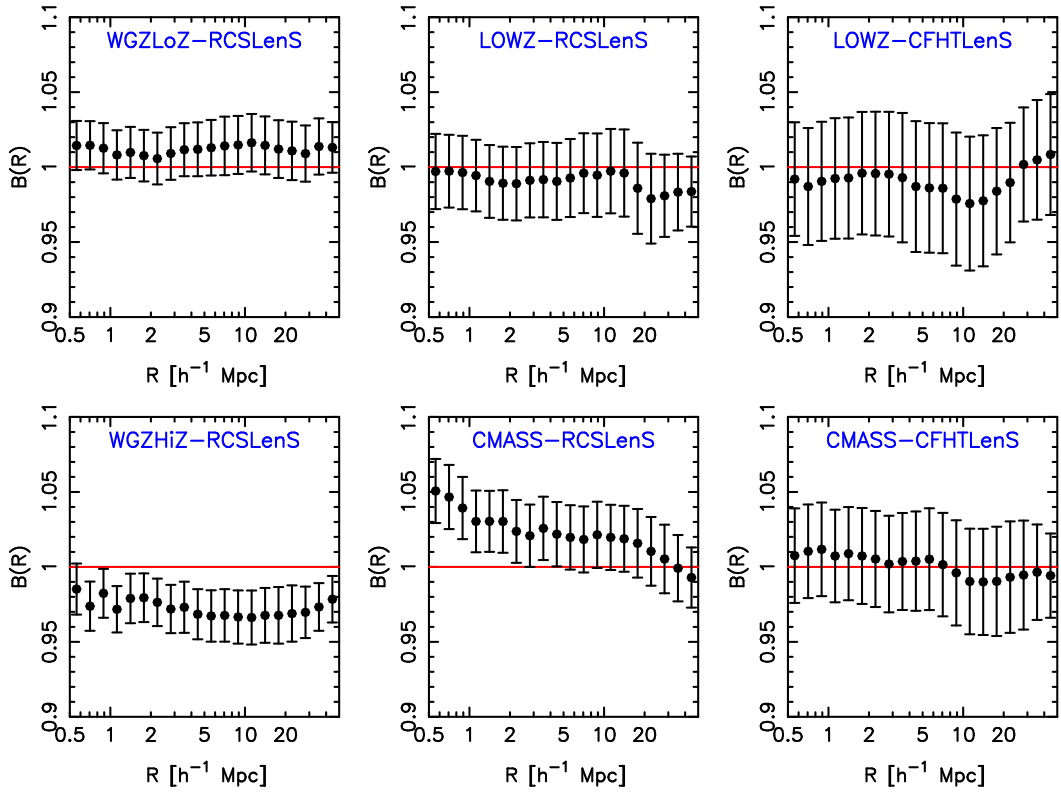}}}
\end{center}
\caption{Boost factor $B(R)$ from source-lens clustering for the
  different combinations of source and lens samples analyzed in this
  study.  The measurements are the average over 40 random catalogues,
  with the errors indicating the standard deviation between the
  catalogues.}
\label{figcorrbr}
\end{figure*}

Figure \ref{figcorrbr} displays the boost factors $B(R)$ for shapes in
the different RCSLenS and CFHTLenS regions correlated with WiggleZ and
BOSS lenses, averaging over 40 random catalogues.  The signal is
generally consistent with $B = 1$, with deviations at the level of
$3\%$ or below for scales $R > 2 \, h^{-1}$ Mpc.  We do not apply this
correction to our measurements.

We note that an effect $B(R) < 1$, which is not expected through
physical association, may be obtained if there is an anti-correlation
between sources and lenses (for example, star-forming galaxies avoid
dense environments on small scales), or through instrumental effects
impacting the imaging data (such as contamination from light of
foreground galaxies or sky subtraction).

\end{document}